\documentclass[reprint,amsmath,amssymb,aps,superscriptaddress
]{revtex4-2}

\usepackage{graphicx}
\usepackage{dcolumn}
\usepackage{bm}
\usepackage{hyperref}
\usepackage{float}
\usepackage{color}
\usepackage{xcolor}
\usepackage{amsmath}
\usepackage{physics}
\usepackage{tikz}
\usetikzlibrary{shapes.geometric}

\begin{document}

\preprint{APS/123-QED}

\title{Quantifying Pauli Errors in Single-Photon Resource-State Generation}

\author{Yuktee Gupta}
\email{yuktee.gupta@sydney.edu.au}
\affiliation{
Centre for Engineered Quantum Systems, School of Physics, The University of Sydney, NSW 2006, Australia
}
\affiliation{Sydney Quantum Academy, Sydney, NSW, Australia}

\author{Andrew C. Doherty}
\affiliation{
Centre for Engineered Quantum Systems, School of Physics, The University of Sydney, NSW 2006, Australia
}

\author{Sahand Mahmoodian}%
\email{sahand.mahmoodian@sydney.edu.au}
\affiliation{
Centre for Engineered Quantum Systems, School of Physics, The University of Sydney, NSW 2006, Australia
}
\affiliation{%
Institute for Photonics and Optical Sciences (IPOS), School of Physics, The University of Sydney, NSW 2006, Australia
}%

\begin{abstract}
We propose a scheme to compute Pauli error rates in photonics-based quantum error correction using experimental observables of single photons produced from quantum emitters. 
We show that first-order coherence measurements and first-order cross correlations---which can be implemented using photon counting---can extract single-photon and entangled single-photon wavefunctions in the presence of imperfections due to photon distinguishability, laser noise, and photon loss. Leveraging this, we show that the wavefunction of any entangled state of noisy photons produced from a single quantum emitter can be expressed in matrix--product-state form and can be used to analytically compute the expectation value of the stabilizer generators of the corresponding entangled state. From this we obtain analytic expressions for the Pauli error probabilities in terms of the photon noise parameters.  Furthermore, we calculate Pauli error maps for entangled photons after undergoing Bell-state measurements in terms of these parameters. Our work provides a method to use experimental measurements to determine the required quality of photons produced from quantum emitters for fault-tolerant fusion-based photonics quantum computing.

\end{abstract}
\maketitle

\section{Introduction}
 
Photonics has emerged as an attractive paradigm for fault-tolerant quantum computing. 
While there are different schemes for photonic quantum computing, the most prominent uses physical qubits encoded as dual-rail single photons. A recent scheme for implementing fault-tolerant quantum computing is fusion-based quantum computing (FBQC) \cite{Bartolucci2021NatComm}. Unlike regular measurement-based quantum computing \cite{Raussendorf2001PhysRevLett, Raussendorf2003PhysRevA},
FBQC requires generating resource states of a fixed size which is independent of the computation and code distance. In FBQC, the photonic circuits performing computation use a combination of linear optics and Bell-state measurements or \textit{fusions}. 

The main challenge in FBQC is producing resource states. Conventional schemes using integrated photonics \cite{Alexander2025Nature} rely on first probabilistically producing single photons using spontaneous four-wave mixing, and then using the single photons to generate three-photon Greenberger-Horn-Zeilinger (GHZ) states probabilistically. The GHZ states are then fused to form the resource state, e.g., `6-ring' states or `loopy diamond' \cite{Bartolucci2021NatComm, Bartolucci2021Arxiv}. This process clearly has significant overhead due to generation schemes being probabilistic at each level and since fusions themselves are inherently probabilistic when using linear optics \cite{Calsamiglia2001AppPhysB}.

Quantum emitters such as atoms, quantum dots, and diamond vacancy centres offer the potential to generate single photons \cite{Sipahigil2014PhysRevLet, Kuhn2002AmPhysSoc, Tomm2021Nature, Zhang2019Light, Hauser2026NatComm, Lee2019IOPPubl, Zhai2022NatureNano, Xing2023arXiv} or even entangled states of photons \cite{Schwartz2016Scien, Cogan2023NatPhot, Thomas2022Nature} deterministically. This is an attractive alternative to resource states generated by repeated fusions \cite{Lindner2009PhysRevA, Paesani2023PhysRevLet}. A quantum emitter with two ground states, and optical transitions for at least one of these states can be used to generate a subset of states representable as Matrix Product States (MPS) with a bond dimension of at-most two \cite{Li2022NPJQuantInf}. This subset includes $n$-photon GHZ states, graph states representable as chains, or chains of GHZ states. Starting with these states, there are methods known to be able to reach any target state of an arbitrary graph with minimal fusions \cite{Lobl2025AmPhysSoc}, or to construct a fusion network as desired.

While quantum emitters potentially enable deterministic state generation, they typically suffer from increased noise, either due to their solid-state environment, e.g., phonons in solid-state systems, or due to experimental imperfections, e.g., unwanted drive lasers polluting single-photon collection channels. These imperfections lead to reduced photon coherence, which can usually be measured by quantum optical tomography such as Hong-Ou-Mandel interference \cite{Hong1987PhysRevLett, Kiraz2004PhysRevA} or, as we consider here, first-order-coherence measurements. Nevertheless there is a tension between quantum optical tomography and quantum computing: quantum optics of propagating photons deals with continuous fields while quantum computing and error-correction protocols are usually based on quantum information encoded as qubits and errors are usually expressed as single-qubit Pauli operations. Although previous work calculated fidelities for cluster/GHZ states \cite{Tiurev2021PhysRevA}, thus far, a rigorous scheme to relate imperfections, as measured in quantum optics experiments, into Pauli error models that can be interfaced with quantum error correction remains missing. 

Here, we develop a new scheme to extract Pauli error models for entangled photons generated from a single quantum emitter. We derive the rate of Pauli errors analytically, purely as a function of experimentally measured parameters, with compact expressions when errors are considered to first order.  We also calculate the additional errors that fusion measurements apply for noisy input states. Existing literature on emitters typically assumes specific sources of noise in emitters \cite{IlesSmith2017NPHOT, Tiurev2021PhysRevA}, while in our model we only consider the influence of the noise on the structure of the photon wavefunction, which accommodates a diverse class of photon noise. On the other hand, most theoretical works in QEC assume purely phenomenological Pauli error models in photonic states \cite{Paesani2023PhysRevLet}, which may not be precise enough to capture the physics of photons used in photonic circuits. We aim to bridge these discrepancies and provide estimates that may inform experimental benchmarks for emitters as well as design for QEC algorithms.

This paper is organised as follows: In Sec. \ref{sec:CohTomography} we derive the structure of a wavefunction for an emitted photon from a general emitter, and then apply this to entangled-photon generation schemes using three-level and four-level emitters. In Sec. \ref{sec:StabExp}, we calculate error probabilities on produced states, using stabiliser expectation values. In Sec. \ref{sec:Fusion} we derive the mapping noisy Bell-states undergo after fusion measurements. Finally, we provide a brief discussion of what our method predicts as possible experimentally allowed error parameters and concluding remarks in Sec.~\ref{sec:discussionAndConclusion}.

\section{First-order-coherence tomography} \label{sec:CohTomography}

The Hong-Ou-Mandel visibility ($\mathcal{V}$) is usually taken as a proxy for the quality of the photons produced by a single-photon source. This is reasonable since it can be used to approximate the wavefunction in the limit $1-\mathcal{V} \ll 1$ \cite{Sparrow2018ICL}.  However this does not allow reconstruction of the single-photon wavefunction in general, which might contain further information which induces errors in the state at a later stage in the photonic circuit. 
 
In this section we show that the first-order coherence function offers a slightly more attractive route to characterise noise in single photon emitters.  We show that this can be used to reconstruct the full state of both single-photon or entangled single-photons (e.g. cluster states) in the presence of indistinguishability errors, coherent laser noise, and loss. Importantly, our approach makes no assumptions about the source of the noise (e.g. phonon sideband vs pure dephasing). We do not consider double-excitation events, considering they can be minimised by using shorter laser pulses, or by using Raman transitions.
 
\begin{figure}
    \centering
    \includegraphics[width=1\linewidth]{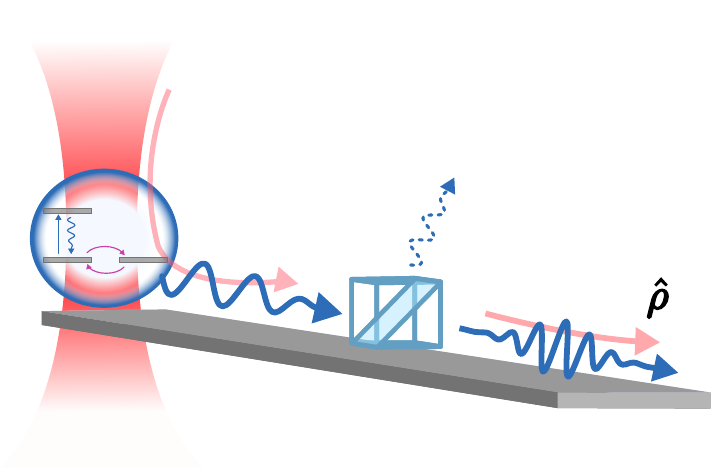}
    \caption{Modelling coherent state leakage and photon loss with beam-splitters. The emitted state is denoted by the solid blue lines and the lost photon is given by the dotted line, after passing through a beamsplitter with transmission coefficient $\eta$. The laser used to excite the emitter can leak into the waveguide, given by the straight red lines, and has final photon number $|\gamma|^2$ which is included in the final density state $\hat{\rho}$}
    \label{fig:bmspltWvgde}
\end{figure}

\subsection{Wavefunctions of single-photons}
We model a noisy quantum emitter being excited by an external laser as shown in Fig.~\ref{fig:bmspltWvgde}. The excited emitter decays and produces a photon. Maintaining generality, we consider the expression for the emitted photon to be a density matrix in the single-excitation sector characterised by the mixed single-photon mode function $f(x,x')$.  Due to fabrication imperfections, a small fraction of the excitation laser can scatter into the collection channel of the quantum emitter. Since the excitation laser pulse has a width that is typically much shorter than the single-photon pulse width, we assume the coherent state is in a temporally orthogonal mode to the single photon. Under this approximation the most general expression of light captured in the collection channel (photon, with some amount of coherent state leakage) is
\begin{equation}
\hat{\rho} = \int dxdx' f(x,x')  \hat{a}^\dagger(x)\ket{0}_{\rm w}\ket{\tilde{\gamma}}_{\rm wc}\bra{\tilde{\gamma}}_{\rm wc}\bra{0}_{\rm w} \hat{a}(x').    
\end{equation}
Here, $\hat{a}(x)$ ($\hat{a}^\dagger(x)$) is the annihilation (creation) operator for a photon at position $x$ in the waveguide channel denoted by the subscript `w'. The subscript `wc' denotes the temporal mode occupied by the coherent state in the waveguide channel. As mentioned before, we assume the coherent state is in an orthogonal temporal mode to the single photon, but in the same spatial mode. Here, $\Tilde{\gamma}$ is the unknown coherent state amplitude of photons that leak into the waveguide. 

The single photon also undergoes two sources of loss. The first being where the photon is emitted into free space, rather than the waveguide. The second occurs during propagation, where the photon gets scattered into the environment due to waveguide imperfections.

The density matrix for the photonic output state is then that of the mixed-state single photon passing through a displacement channel, and then through a vacuum loss channel. We model the output density matrix as containing the emitted photon as well as leaked laser light transmitted through a virtual beamsplitter with a transmission coefficient $\eta$ (see Fig.~\ref{fig:bmspltWvgde}), 
\begin{align}
&\hat{\rho} = (1-\eta)\ket{0}_{\rm w}\ket{\gamma}_{\rm wc}\bra{\gamma}_{\rm wc}\bra{0}_{\rm w} \nonumber \\
&+ \int dxdx' ~ f(x,x')  ~  \hat{a}^\dagger(x)\ket{0}_{\rm w}\ket{\gamma}_{\rm wc}\bra{\gamma}_{\rm wc}\bra{0}_{\rm w} \hat{a}(x'),
\end{align}
where $\gamma = \sqrt{\eta}\Tilde{\gamma}$, the remaining coherent state amplitude. Our aim is now to determine a scheme to quantify the unknown parameters $\eta, \gamma$ and $f(x,x')$. These can be obtained by performing tomography on this state. Measuring the unnormalised first-order coherence function of the output gives, 
\begin{align}
G^{(1)}(x_1,x_2) &= \langle\hat{a}^\dagger(x_2)\hat{a}(x_1)\rangle
\nonumber \\
&= \eta f(x_1,x_2) +|\gamma|^2c(x_1)c^*(x_2).
\end{align}
This is experimentally measurable using a standard Mach-Zehnder Interferometer, with a delay in one arm, and photon counting at the outputs \cite{Wang2025NCOMM}. Alternatively, it can be measured using balanced homodyning of the output~\cite{Qin2015Nature}. Here, $c(x)$ is the normalised temporal mode of the laser in the waveguide.  We assume that $c(x)$, and the average photon number of the coherent field that leaks in i.e. $|\gamma|^2$, are known as they can be determined by measuring the excitation field in the absence of any photon from the emitter. 

Using the fact that $f(x,x')$ has unit trace, we can obtain,
\begin{equation}
\eta = \int dx ~G^{(1)}(x,x) - |\gamma|^2.
\end{equation}
With this, $f(x,x')$ is now known from the expression of $G^{(1)}(x,x')$.  Further, we can diagonalise $f(x,x')$
since it forms the reduced density matrix of the single photons after tracing out the coherent state, and thus is Hermitian and positive semidefinite. Its eigenvectors give us a complete set of basis modes that span the Hilbert space of the emitted photon, and the eigenvalues give us the average photon number in each mode. We can then express $\hat{\rho}$ in this basis:
\begin{align}
\hat{\rho} = &\left( (1-\eta)\ket{0}_{\rm w}\bra{0}_{\rm w} + \eta \sum_n |\alpha_n|^2\ket{1}_n\bra{1}_n \right) \nonumber\\
&\otimes \ket{\gamma}_{\rm wc}\bra{\gamma}_{\rm wc}
\end{align}
Here, $\ket{1}_n = \int dx ~ v_n(x) \hat{a}(x) \ket{0}$ , where $v_n(x)$ are the orthogonal eigenvectors/eigenfunctions of $f(x,x')$ after diagonalization, and $|\alpha_n|^2$ are the eigenvalues. Since there is only a single excitation emitted, we also have $\sum_n |\alpha_n|^2 = 1$.

As we will show in the coming sections, for the purpose of calculating stabiliser expectation values for entangled photon states, it is convenient to express the state of the photon using a pure state. Given that we have diagonalised $\hat{\rho}$, we can equivalently express the photon state as a pure state, with a set of labelled auxiliary states $\ket{\phi_n}$, which we attribute to the environment becoming entangled with the emitter. The exact form of $\ket{\phi_n}$ is not necessary to be known, we only utilise the fact it is orthogonal to other labelled environmental states, $\langle \phi_m | \phi_n \rangle = \delta_{mn}$. We can thus write the pure state as,
\begin{equation}\label{phi_init}
\ket{\psi} =  \sum_n \alpha_n\ket{\phi_n} \ket{1^{\rm ph}_n}\ket{\gamma}_{\rm wc},    
\end{equation}
where $\ket{1^{\rm ph}_n} = \left[ \sqrt{\eta}\ket{1_n}_{\rm w}\ket{0}_{\rm l} + \sqrt{1-\eta}\ket{0}_{\rm w}\ket{1_n}_{\rm l}  \right]$. Here the subscript `$\rm l$' is to denote the loss channel, as opposed to `$\rm w$', the waveguide channel. Tracing out the environmental state and the loss channel, we get back $\hat{\rho}$ as before. We now have a general expression for an emitted photon, where all parameters of the wavefunction can be determined by measuring the first-order coherence function of the photon source.

\subsection{Wavefunctions of entangled single photons}\label{sec:Wavefuncions}

Two main schemes for encoding the qubits as photons using quantum emitters and producing entangled states have been previously proposed \cite{Lindner2009PhysRevA, Tiurev2021PhysRevA}. These works use  a three-level emitter to implement a time-bin encoding or a four-level emitter to implement a polarisation encoding. As we show in this paper, the two distinct schemes have different error profiles. For quantum computing proposals, quantum emitters are used to deterministically produce what we call `caterpillar states' with each physical qubit encoded as a single photon \cite{Paesani2023PhysRevLet}.  These states are unitarily equivalent to graph states represented by caterpillar trees/graphs~\cite{West2001PrenticeHall} where any 'branched' vertex is a maximum of one link away from the central path in the graph. They are analysed separately in the following subsections.
Continuing the approach of the previous subsection, we show how the first-order coherence function can be used to extract the parameters determining the imperfections of the now entangled multi-photon state, and obtain an analytic expression of the entire emitted wavefunction, which can be compactly expressed as a bond-dimension-two MPS.

\begin{figure}
    \centering
    \includegraphics[width=\columnwidth]{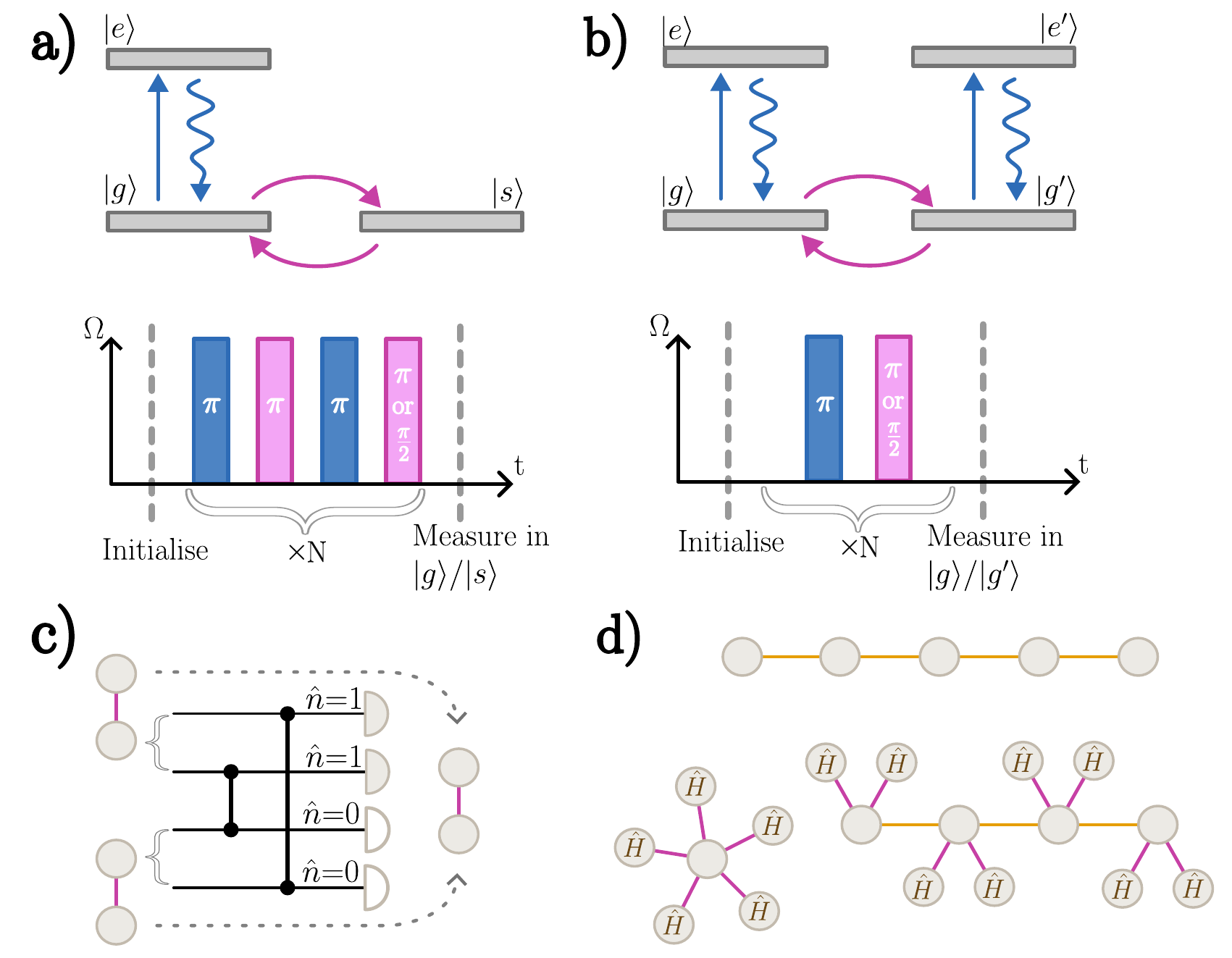}
    \caption{Level schemes of emitters and applied pulse sequences for the corresponding emitters. (a) The three-level emitter with two degenerate ground states, and the excited state is only accessible by one of the ground states, $\ket{g}$. One pulse sequence here involves exciting the emitter (blue bar), performing a $\pi_{s\leftrightarrow g}$ rotation (pink bar), exciting the emitter again and then performing a final $\pi_{s\leftrightarrow g}$/$\frac{\pi}{2}_{s\leftrightarrow g}$ rotation to emit a GHZ/cluster state. (b) The four-level emitter with two degenerate ground/excited states. One pulse sequence here involves exciting the emitter (blue bar) and emitting and then performing a $\pi_{g\leftrightarrow g'}$/$\frac{\pi}{2}_{g\leftrightarrow g'}$ pulse (pink bar) for the GHZ/cluster state. (c) The dual-rail fusion circuit we use here showing the input of one photon each from two Bell states, $\ket{\Phi^+}$, and detecting $\ket{1100}$/$\ket{0011}$/$\ket{1010}$/$\ket{0101}$ the resultant output is a $\ket{\Phi^\pm}$ Bell state with probability 1/2.  (d) Examples of some states emitted: GHZ states, chains of photons, or branched chains}
    \label{fig:levelsOverall}
\end{figure}

\subsubsection{Three-Level Emitter}\label{subsec:Wavefuncions3}

Consider a quantum emitter with an arrangement of three energy levels with a single optical transition, as proposed in \cite{Tiurev2021PhysRevA}, shown in  Fig.~\ref{fig:levelsOverall}(a).  The photon is produced when the emitter decays from the excited state $\ket{e}$ to ground state $\ket{g}$. The third level is a degenerate ground state $\ket{s}$ only accessible from $\ket{g}$ by a separate laser or microwave source. Initialising the emitter to $\ket{g}$ or $\ket{s}$ and exciting it, or applying a $\pi$-pulse and then exciting it, produces an early or a late photon. The early and late photons are temporally orthogonal, and serve as our qubit basis,  $\ket{0}$ and $\ket{1}$ respectively, more commonly known as a time-bin encoding. With the help of a switch this encoding can be converted to and from a dual-rail encoding. We assume one can switch between both encodings ideally, so as to focus on the emitter induced noise. Full-scale simulations would however need to take into account noise and loss added by the non-linear switches that would be required to convert between the encodings.
As shown in Fig.~\ref{fig:levelsOverall}(a), every qubit emitted consists of two laser excitation rounds to account for early or late photons. We express the photon states using the expression of an emitted photon as in \eqref{phi_init}, with the additional environmental state and coherent leakage for the round of laser excitation that doesn't emit a photon. Using \eqref{phi_init}, a photon emission process maps $\ket{e}\otimes \ket{0} \rightarrow \ket{g}\otimes\ket{\psi}$. We then have the following expressions for the qubits,
\begin{align}\label{01tildeDef}
\ket{\tilde{0}}&= \left(\sum_n \alpha_n \ket{\phi_n}\ket{1^{\rm ph}_n}\ket{\gamma}_{\rm wc} \right) \otimes \ket{\phi_s}\ket{0}\ket{\gamma}_{\rm wc}, \\
\ket{\tilde{1}} &= \ket{\phi_s}\ket{0}\ket{\gamma}_{\rm wc} \otimes \left(\sum_n \alpha_n \ket{\phi_n}\ket{1^{\rm ph}_n}\ket{\gamma}_{\rm wc} \right).
\end{align}
The $\ket{\phi_n}$ here are the orthonormal environmental states associated with the photon emission from the excited state $\ket{e}$ as in \eqref{phi_init}, and $\ket{\phi_s}$ is the environmental state associated with $\ket{s}$. This is the state of the environment when the emitter is in the $\ket{s}$ state and does not emit a photon. We can express $\ket{\phi_s} = \sum_n \beta_n \ket{\phi_n} + \delta \ket{\phi_{\perp}}$  for $\beta_n = \braket{\phi_n}{\phi_s}$ and where $\ket{\phi_\perp}$ is the remaining environmental state that is in a space orthogonal to all $\ket{\phi_n}$ with amplitude $\delta$.

The photon generation protocols for GHZ and entangled chains are as shown in Fig.~\ref{fig:levelsOverall}(a).  Preparing the emitter in a superposition of $\ket{g}$ and $\ket{s}$, repeatedly exciting the emitter and then performing a $\pi_{s\leftrightarrow g}$ rotation, produces an $n$-photon GHZ state of $\ket{\tilde{0}}$ and $\ket{\tilde{1}}$. An alternate pulse sequence replacing every second $\pi_{s\leftrightarrow g}$ rotation by a Hadamard gate $\hat{H}_{s \leftrightarrow g}$, produces a chain of photons in an entangled cluster state . Any combination of these two pulse sequences produces caterpillar states (for example branched chains~\cite{Buterakos2017PhysRevX}).

To determine the $\beta_n$, we can use a Mach-Zehnder with a relative delay of the two arms to account for the time difference between the early and late photons to measure the first-order coherence function as a cross correlation. Consider the correlator $G^{(1)}(t_1,t_2) = \langle \hat{a}^\dagger(t_1)\hat{a}(t_2)\rangle$, where we consider times $t_2 \sim t_1 +\tau $, where $\tau$ is the delay between early and late photons. This means that $t_2$ is delayed to account for the time it takes to emit the second photon. The observable is therefore a first-order cross correlation of the emitted photons. For a the single photon emitted in an equal superposition of early and late $\ket{\psi} = \frac{1}{\sqrt{2}} \sum_n \alpha_n  ~ \left(~\ket{\phi_n\phi_s}\ket{1_n0}+ \ket{\phi_s\phi_n}\ket{01_n}~\right)$, we have:
\begin{align}
    G^{(1)}(t_1,t_2) = \frac{1}{2}\sum_{x,y}  &\alpha_x^*\alpha_y v_x^*(t_1) v_y(t_2) \nonumber\\ 
    \times &\left[\eta\beta_x\beta_y^*(1+|\gamma|^2)^2 + 2(1-\eta)|\gamma|^2 \right],
\end{align}
Using the orthogonality of the $v_n(t)$, we can obtain the norm values of the different $\beta_n$, 
\begin{align}
    |\beta_n|^2 = &\frac{\left(2 \iint_{t_1, t_2} v_n(t_1) v_n^*(t_2) G^{(1)}(t_1, t_2) dt_1 dt_2 \right)}{\eta~(1+|\gamma|^2)^2|\alpha_n|^2} \nonumber\\
    &- (1-\eta)|\gamma|^2.
\end{align}
The $\alpha_n$ and functions $v_n(t)$ are known as in \eqref{phi_init} from performing single-photon tomography. Subsequently the value of $\delta$ is known if we know all $|\beta_n|^2$ values,  $|\delta|^2 = 1 - \sum_n |\beta_n|^2$. Note that knowing the absolute values $|\beta_n|, |\delta|$ is sufficient for our noise calculations and the the phase does not play a role. A detailed calculation is presented in Appendix~\ref{BetaDeriv}. We see in later sections that the relevant quantity for the interference of two photons is a function of the $\beta_n$ and $\alpha_n$, what we call  $\zeta = \sum_n |\alpha_n|^2|\beta_n|^2$.  We interpret $\zeta^2$ as  being akin to a weighted visibility-like term, which reflects environmental entanglement.
We thus conclude that even with the addition of the auxiliary environmental states, the correlator $G^{(1)}$ allows us to determine the complete theoretical expression of resource states emitted in a three-level emitter.

\subsubsection{Four-Level Emitter}\label{subsec:Wavefuncions4}

The second level scheme we consider is a four-level emitter with optical transitions having orthogonal transition-dipole moments each linked to a different ground state. This level scheme can also be sequentially pulsed to produce entangled photons as previously reported \cite{Lindner2009PhysRevA, Cogan2023NatPhot, Wilk2007Science}. This is shown in Fig.~\ref{fig:levelsOverall}(b). Here the two degenerate ground states are $\ket{g}, \ket{g'}$ as well as degenerate excited states $\ket{e}, \ket{e'}$.  Photons are produced by the two emission process, each with the  mapping  $\ket{e}\otimes\ket{0} \rightarrow \ket{g}\otimes \ket{L}$ or  $\ket{e'}\otimes\ket{0} \rightarrow \ket{g'}\otimes \ket{R}$. Each of the emissions are coupled to different waveguide modes. Ideally, the states $\ket{R}$ and $\ket{L}$ are emitted into orthogonal modes and serve as our qubit basis. In the original work \cite{Lindner2009PhysRevA}, the qubit was proposed to be encoded as right- and left-hand circularly polarised waves, however this can also be realised using a polarisation-degenerate cavity \cite{Antoniadis2022NPJQuantInf}, or on-chip implementations which can couple right- and left-hand circularly polarised transition dipoles to right- and left-propagating fields using chiral light-matter interaction \cite{Lodahl2017Nature, Sollner2015NNANO} and therefore the propagation direction can encode the qubit as well. Irrespective of the scheme, in practical implementations, the emitter has a slight birefringence and each transition has some parasitic coupling to the mode to which it should otherwise be orthogonal. We quantify this with a directionality parameter $\mathcal{D}$, which is ideally 1. This results in emitted states that are not orthogonal and have finite overlap dependent on $\mathcal{D}$. We denote the states emitted from the emitter as $\ket{L}/\ket{R}$, and the final states in the waveguide as $\ket{\bar{0}}/\ket{\bar{1}}$.
\begin{align}
    \ket{L} &= \sum_n \alpha_n\ket{\phi_n}\ket{1^{\rm ph}_n}_{\rm L}
    \ket{\frac{\gamma}{\sqrt{2}}}_{\rm Rc}\ket{\frac{\gamma}{\sqrt{2}}}_{\rm Lc},\\
    \ket{R} &= \sum_n \alpha_n\ket{\phi_n}\ket{1^{\rm ph}_n}_{\rm R} ~
    \ket{\frac{\gamma}{\sqrt{2}}}_{\rm Rc}\ket{\frac{\gamma}{\sqrt{2}}}_{\rm Lc},
\end{align}
and
\begin{align}\label{01barDef}
    \ket{\bar{0}} = \sqrt{\mathcal{D}}~\ket{L} + \sqrt{1-\mathcal{D}}~\ket{R}, \\
    \ket{\bar{1}} = \sqrt{\mathcal{D}}~\ket{R} + \sqrt{1-\mathcal{D}}~\ket{L}
\end{align}
We have $\rm L/R$  subscripts for the left/right waveguides, $\rm Rc/Lc$ for the coherent states in the respective waveguides. Note the loss from the respective waveguides is considered orthogonal as well ($\rm Rl/Ll$  ).

Initialising the emitter in a superposition of $\ket{g}, \ket{g'}$, repeatedly exciting the state and emitting photons produces a $N$-GHZ state. Performing a ${\frac{\pi}{2}}_{g \leftrightarrow g'}$ pulse after every emission pulse produces an entangled chain cluster state. Since the excited states typically only differ due to their spins, we assume that the environment couples to these states equally and does not become entangled with the internal states of the emitter. We can see that as $\ket{\phi_n}$ factors out similarly for $\ket{\bar{0}}$ and $\ket{\bar{1}}$.

\subsection{Matrix-product-state encoding of noisy entangled photons}\label{sec:mpsnoisy}

We now have a scheme to use experimentally measured parameters to obtain full expressions of the emitted graph sates. Calculating any expectation values on these expressions is however tedious when dealing with a large number of photons. Here we can note that the noise considered in each step of excitation is local to the photon emitted (assuming a full de-excitation of the emitter every time), and not correlated to photons emitted before or after. This means that the noise does not introduce any additional entanglement into the system. Since the protocols use schemes with two ground states, the entangled photon states generated can be represented by MPSs with at most bond dimension two\cite{Li2022NPJQuantInf}. As we show, this allows us to write down a compact expression for emitted graph states and  calculate expectation values of operators in a straightforward way. The GHZ and cluster states can simply be written as,
\begin{align} \label{eq:matrix_repr}
\text{$N$-GHZ :} ~~~~~~~ \frac{1}{\sqrt{2}}(\ket{0000 ....} + \ket{1111 ....})\nonumber \\
= 
\frac{1}{\sqrt{2}}
~
\begin{bmatrix}
    1 & 1 \\
\end{bmatrix}
\begin{bmatrix}
    \ket{0} & 0 \\
    0 & \ket{1}
\end{bmatrix} ^{\times N}
\begin{bmatrix}
    1 \\
    1
\end{bmatrix},\\
\text{$N$-long chain} : \sum_{b \in \{0,1\}^{\otimes N}} (-1)^{f(b)}\ket{b} \nonumber
\\
\frac{1}{\sqrt{2}^N}
\begin{bmatrix}
    1 & 1 \\
\end{bmatrix}
\begin{bmatrix}
    \ket{0} & \ket{0} \\
    \ket{1} & -\ket{1}
\end{bmatrix} ^{\times N}
\begin{bmatrix}
    1 \\
    0
\end{bmatrix}.
\end{align}
Here $\ket{0}/\ket{1}$ refer to the noisy physical photonic qubit states emitted in the waveguide channel corresponding to  $\ket{\tilde{0}}/\ket{\tilde{1}}$, defined in \eqref{01tildeDef},  or $\ket{\bar{0}}/\ket{\bar{1}}$, defined in \eqref{01barDef}, emitted by the three and four-level emitters respectively, Finally,  we note that $b$ is any $N$-digit binary number, and $f(b)$ is  a function of $b$, such that it is equal to zero (one) for an even (odd) number of nearest neighbour pairs of 1 in $b$.

We have described the pulse sequences to produce GHZ and chain states, in three and four level emitters. These pulse sequences can also be interleaved to create branched chains as seen in Fig.\ref{fig:levelsOverall}(d), or generally caterpillar states\cite{Buterakos2017PhysRevX}. These branched chain states are similarly representable as bond-dimension-two MPS with a layout shown schematically in Fig.~\ref{fig:MPSdia}. Further graph states might be accessible by local complementation or single qubit measurements~\cite{Lobl2025AmPhysSoc, Hein2004PhysRevA}, giving us access to a large number of cluster states with a single emitter.\\
We now have a full wave-function expression of the output of an emitter, as well as a convenient representation in terms of MPSs. Diagrammatically we can show MPSs using the Penrose notation (see Fig. \ref{fig:MPSdia}), where each matrix in Eq. \ref{eq:matrix_repr}, corresponding to each photonic qubit, is represented by a square. The horizontal legs represent the dimension of the matrices, the so called  bond dimension of the MPS. The vertical leg accesses the possible states each photon can be. See Ref.~\cite{Schollwock2011Annals} for a detailed review on MPSs.

Many works model photons as being in a superposition of an ideal and an orthogonal `bad' state \cite{Sparrow2018ICL, Kiraz2004PhysRevA}, $\ket{\psi_i} = \mathcal{V}^{\frac{1}{4}}\ket{1_{\rm ideal}} + {(1-\sqrt{\mathcal{V}})}^{\frac{1}{2}}\ket{1_{i}}$  where the state is labelled by the non-ideal state $\ket{1_i}$, orthogonal to the ideal state and other non-ideal states $\langle 1_j|1_i \rangle =0, ~~ \forall ~ i\neq j$. This ensures the Hong-Ou-Mandel visibility of any photon pair is  $|\langle\psi_j |\psi_i\rangle|^2 = |\mathcal{V}^{\frac{1}{2}}\langle 1_{\rm ideal}|1_{\rm ideal}\rangle + (1-\sqrt{\mathcal{V}})\langle 1_{j}|1_{i}\rangle|^2 = \mathcal{V}$. This model may not be able to capture effects of multiple temporal modes completely, which becomes more important in the limit of larger noise. This single-mode approximation is unnecessary when the full spectral wavefunction is known. In this work we use the complete wave-functions of emitted states, expressed as MPS in terms of experimentally obtained parameters. This allows us to capture effects of noise, without approximation, on error rates relevant to quantum error correction thresholds, which we derive in the next section.

\section{Calculating Stabiliser Expectation Values and Error Probabilities} \label{sec:StabExp}

\begin{figure}
    \centering
    \includegraphics[width=\columnwidth]{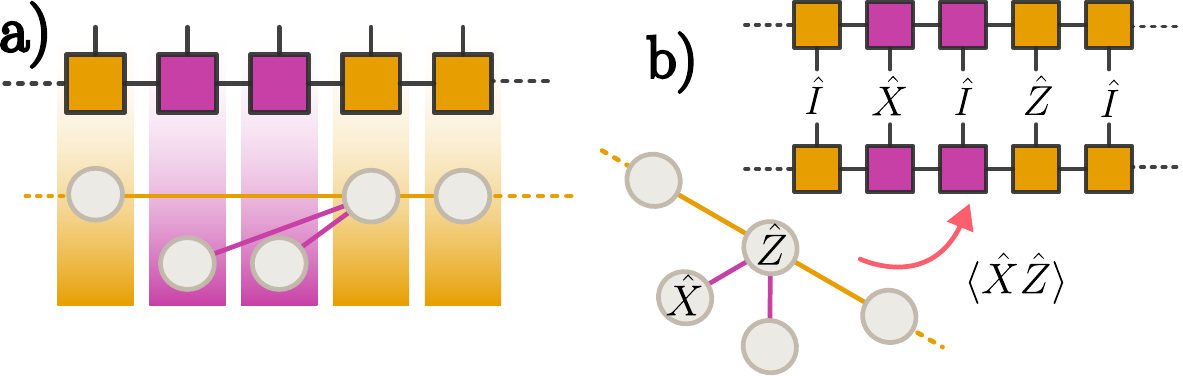}
    \caption{Representing emitted photons as an MPS, where we have used the Penrose graphical notation to represent the MPS. (a) shows the graphical MPS representation of the state vector of a graph state below. GHZ photons and cluster state photons represented by different types of MPS blocks as in Eq.~\ref{eq:matrix_repr}, here represented in orange and pink, b) Calculating the expectation values of an operator on the graph state is simplified in this representation. The expectation value is simply the MPO of the operator contracted with the MPS representation of the state, where operators act on corresponding MPS blocks.}
    \label{fig:MPSdia}
\end{figure}

To determine whether the noisy entangled states in Sec.\ref{sec:Wavefuncions} are of sufficient quality to use for error-corrected quantum computation, it is necessary to quantify the quality of the produced states. One way to do this is by measuring the fidelity of the states, with respect to an ideal state. There are two reasons why the fidelity of the state is not a sufficiently complete description of its noise. Firstly, in terms of the defined $\ket{\bar{0}/\bar{1}}$ or $\ket{\tilde{0}/\tilde{1}}$ as in equations \eqref{01barDef} and \eqref{01tildeDef}, the ideal states would have some $\alpha_j = 1$ and $\alpha_k =0 ~ \forall ~ k\neq j$. However the choice of mode $j$ is not obvious. We can set $j$ to be the mode most probabilistically produced, i.e. largest  $|\alpha_j|^2$, but this does not completely capture how the quantum information is encoded. Since photonic qubits are measured by number-resolving single-photon detectors, correct detection of a photon does not require it to be in a specific temporal mode, rather it needs to only be in the correct rail and arrive within a specific time cycle. This means that non-ideal modes have the same detection statistics as the ideal mode, on being passed through photonic circuits.  Secondly, even in a single-mode photonic qubit, fidelity tells you how close a state is to a target state. It does not tell you the nature of the noise. In quantum error correction, knowing the nature of the noise is critical. For example, biased Pauli noise of a specific type, is much easier to correct than depolarising noise\cite{Panos2008PRA, Tuckett2019PRX}. Therefore careful characterisation of the noise is necessary.

Here we propose calculating the Pauli error probabilities on the states themselves. We do this by equating the probabilities of measurement outcomes on the photonic graph state, with measurement statistics of corresponding pauli measurements on a noisy $N$-qubit graph state. We characterise the noise in our photonic state with  $p_x, p_y$, and $p_z$ for Pauli $\hat{X}$, $\hat{Y}$, and $\hat{Z}$ errors respectively. We can assume the errors on the states are Pauli, since any coherent error can be mitigated by the use of techniques, such as Pauli twirling\cite{Knill2008PhysRevA}, or randomised compiling \cite{Wallman2016PhysRevA}.

To proceed, we take advantage of the stabiliser formalism. The stabiliser group under multiplication, $\mathsf{S}$ for the ideal $N$-qubit cluster state,  is a set of  $2^N$ Pauli operators, with $N$ generators, of which the ideal state $\ket{\psi}_{\rm ideal}$ is a simultaneous $+1$ eigenstate, $\forall S_j \in \mathsf{S} ~;~ S_j\ket{\psi}_{\rm ideal} = \ket{\psi}_{\rm ideal}$. The generating set is a subset of $N$ operators that generate the stabiliser group $\mathsf{S}$. For graph states, this generating set admits a graphical representation. The graph state is a associated with a graph $G\in \{ V, E\}$, with a set of vertices $v_i\in V$ and edges $E$ which is a set of pairs of vertices $\{ v_i, v_j \}$. For every vertex $v_i$, we have a corresponding stabiliser generator : $\mathcal{G}_i = X_i \prod_{\langle i,j\rangle} Z_j$ where $\hat{X}_i$,$\hat{Z}_j$, are the Pauli matrices on sites $i,j$, and $\langle i,j\rangle$ represents that  $\{v_i,v_j\}\in E$. The stabiliser group is then products of these generators, $S_j = \prod_{i=1}^{N} \mathcal{G}_i^{b_i}$ for a binary string $b\in \{0,1\}^{ N}$  and $i^{th}$ digit $b_i$. We should also note the stabilisers can be expressed as a product of Pauli operators acting on each qubit $S_j = P_1 \otimes P_2 \dots P_N$ where $P_i \in \{\hat{I},\hat{X},\hat{Y},\hat{Z}\}$. 

The $\ket{\bar{0} / \bar{1}} ~ \text{or} ~ \ket{\tilde{0}/\tilde{1}}$ states are qubit states encoded in a larger photonic Hilbert space that does not just include the computational state. We would like to then construct operators that act on this larger Hilbert space that act like Pauli operators on the encoded qubits. If we were able to produce the ideal qubit graph state , these constructed operators will have the same detection outcomes as Pauli operators on the ideal state. These operators obey the same commutation rules as Pauli operators. Since these operators correspond to actual measurements we would also like the operators to act on all modes equivalently. That is single-photon counting is not mode sensitive, and as long as a photon arrives, its temporal mode does not influence the measurement outcome. We assume the photon detectors are number resolving and therefore operators must act in the dual-rail photon subspace, and map non-computational multi-photon or vacuum states to zero, since multi-photon and loss errors can be heralded. This means that the constructed operators don't necessarily preserve the trace of the density operator they act on, i.e., the operators are isometries rather than unitary matrices. With these constraints we construct the following operators, denoted by $\hat{I},\hat{X},\hat{Y} \text{~and~} \hat{Z}$ (see Appendix \ref{OpModeDeriv} for derivation):
\begin{align}
\hat{I} &= \sum_n \ket{1_n0}\bra{1_n0} + \ket{01_n}\bra{01_n}, \label{eq:SinglePhotonOps}\\
\hat{X} &= \sum_n \ket{1_n0}\bra{01_n} + \ket{01_n}\bra{1_n0},\label{eq:SinglePhotonOps2}\\
\hat{Y} &= i \sum_n \ket{1_n0}\bra{01_n} - \ket{01_n}\bra{1_n0} ,\label{eq:SinglePhotonOps3}\\
\hat{Z} &= \sum_n \ket{1_n0}\bra{1_n0} - \ket{01_n}\bra{01_n}.\label{eq:SinglePhotonOps4}
\end{align}
Here $n$ are the various photonic modes we sum over, as defined in (\ref{phi_init}).

It is convenient to use the matrix-product operator(MPO) representation of stabilisers as we have the state expressions in MPS form. Every stabiliser is tensor product of operators, acting on a single site of the MPS, hence there is a simple MPO representation of the operators. We can also represent the MPO in terms the $N$-qubit computational basis states $\sigma_{1/2}$: $S_j= \sum_{\sigma_1, \sigma_2 = 0}^{2^n - 1} P_{1,j}^{\sigma_1, \sigma_2} P_{2,j}^{\sigma_1, \sigma_2}... \ket{\sigma_1}\bra{\sigma_2}$ where $P_{i,j}$ is the Pauli operator acting on site $i$. Calculating stabiliser expectation values then is a contraction of the tensor network consisting of the state MPS and stabiliser's MPO as shown schematically in Fig.~\ref{fig:MPSdia}(b).

The stabiliser expectation values then simplify to be functions of single-photon operator expectation values, which in turn are functions of the model parameters we obtained in Sec. \ref{sec:CohTomography}. For example, the Bell state $\ket{\Phi^+} = \frac{\ket{00} + \ket{11}}{\sqrt{2}}$  with stabilisers $\langle \hat{I}\hat{I}\rangle, \langle \hat{X}\hat{X}\rangle, \langle \hat{Z}\hat{Z}\rangle, \langle -\hat{Y}\hat{Y}\rangle$, for the three-level scheme gives us,
\begin{align}
    \langle \hat{I}\hat{I}\rangle &= e^{-4|\gamma|^2} ~ \left({\eta} + {(1-\eta)|\gamma|^2} \right)^2,\\
    \langle \hat{X}\hat{X}\rangle &= e^{-4|\gamma|^2} ~ \left({\zeta^2\eta^2} + {(1-\eta)^2|\gamma|^4} \right),\\
    \langle -\hat{Y}\hat{Y}\rangle &=e^{-4|\gamma|^2} ~ {\zeta^2\eta^2},\\
    \langle \hat{Z}\hat{Z}\rangle &= e^{-4|\gamma|^2} ~ {\eta^2}.   
\end{align}
The exponential prefactor arises from the coherent state drive leaking into the waveguide and taking the photonic state in the waveguide out of the single-excitation subspace as the coherent amplitude increases. The expectation value of the identity operator, here $\langle \hat{I}\hat{I}\rangle$, is not trivially 1, referring to Eq. \ref{eq:SinglePhotonOps}, each $\hat{I}$ counts the probability of detecting a single photon, emitted or leaked in from the excitation laser, in either rail. The expectation value, $\langle\hat{I}^{\otimes N}\rangle$ generally,  then is the total probability of detecting $N$ single-photon states. Since we would like to know the stabiliser expectation values, given we are in the code-space, we normalise all stabiliser expectation values by $\langle\hat{I}^{\otimes N}\rangle$.
Stabiliser expectation values, calculated similarly for the cluster state,  and GHZ state are listed in Appendix \ref{StabVals}.

\subsection{Estimation of Pauli error probabilities from stabiliser expectation values}\label{subsec:PauliErrFromStab}

We can now, in principle, calculate the expectation values of the $2^N$ stabilisers in terms of the model parameters analytically. We now would like to relate these to the error probabilities on the qubits.
All possible $4^N$ Pauli operators that can act on the $N$-qubit state form a group under multiplication, $\mathsf{P} ~;~ P_i \in \{\hat{I},\hat{X},\hat{Y},\hat{Z}\}^{\otimes N}$, of which the stabilisers are a subgroup $\mathsf{S} \subset \mathsf{P}$.  The stabiliser group $\mathsf{S}$ divides the Pauli group $\mathsf{P}$ into $|\mathsf{P}|/|\mathsf{S}| = 2^N$ cosets of size $2^N$ each. Elements within this coset cannot be distinguished, as their effect on the state is identical under application of stabilisers~\cite{Gottesman1997ArXiv}. We would then want to choose the most physically motivated set of errors, which is usually errors with the lowest weight. With these constraints, we choose an $E_i$ from each of these cosets and construct the set of errors $\mathsf{E}$  that may occur on a state.

Using commutation rules of the Pauli group, and the cyclic property of trace we can show that the expectation value of any stabiliser is expressible as a sum of the probabilities of these errors $E_i$.  We have $\hat{\rho} = \sum_i p_i E_i \hat{\rho}_{\rm ideal} E_i^{\dagger}$ where $p_i$ is the probability of error $E_i$ occurring, and $\hat{\rho}_{\rm ideal}$ is the ideal density matrix. Then for any stabiliser,
\begin{align}\label{eq:stabexp_to_pauli}
	\langle S_j \rangle &= \operatorname{Tr}{( S_j \hat{\rho})}
	=  \sum_{i=0}^{2^N}  p_i \operatorname{Tr}{(S_j E_i\hat{\rho}_{\rm ideal} E_i^\dagger )} \\
    &= \sum_{i=0}^{2^N} p_i A_{ij} 
    \text{  ; where }  A_{ij} =
    \begin{cases}
         ~~~1  &\text{ if } ~~ [S_j, E_i] = 0  \\
        -1       & \text{ if } ~~ \{S_j, E_i\} = 0.  
    \end{cases}
\end{align}
This is true for all $2^N$ stabilisers, summarised as the vector equation $\langle \vec{S}\rangle = A\cdot\vec{p}$, where $\langle\vec{S}\rangle$, and $\vec{p}$ are vectors of the Stabiliser expectation values, and error probabilities respectively.  Here, $A$ encodes commutation information, where $A_{ij} =1$ for commuting or $A_{ij} =-1$ for anti-commuting elements. Hence with the stabiliser expectation values calculated, we have the error probabilities $\vec{p} = A^{-1}\cdot \vec{S}$ for all $2^N$ errors. 

Estimating individual errors per qubit from these $2^N$ values without performing exponentially many calculations requires making some assumptions about the structure of the error model. In the models considered here, the emitter de-excites completely after every emission cycle, the photon emitted at steps $i$ and $i+1$  are uncorrelated. This assumption is also built in to how we write out the MPS expression of the state and the fact that it has a bond dimension of two \cite{Li2022NPJQuantInf}, as  correlated errors would lead to a larger bond dimension. We confirm this holds for this model, where log-linear methods applied to $\vec{p}$ show vanishing correlated errors for longer chains. This means that the errors that occur on each emitted photonic qubit are independent. This reduces the number of parameters we need to estimate, to $\approx 3N$, corresponding to $p_x, p_y, p_z$ for each qubit. The number of stabiliser expectation values needed to be known is now $3N$. 

To obtain a closed set of equations for a finite sized state, we need to examine the behaviour of errors on the boundary qubits. For a finite $N$-qubit chain, the stabiliser generators at the boundaries are $\mathcal{G}_1 = \hat{X}_1 \hat{Z}_2$ and $\mathcal{G}_N = \hat{X}_N \hat{Z}_{N-1}$. Recall we only consider the unique errors $E_i \in \mathsf{E}$, an $\hat{X}$ error on qubit 1 is equivalent to a $\hat{Z}$ error on qubit 2 under the application of $\mathcal{G}_1$, hence we cannot estimate probabilities for both independently. Similarly, we cannot estimate both $p_{x,N}$ and $p_{z,N-1}$. The number of errors we need to estimate for a chain is then $3N-2$, with a choice of discarding $\hat{X}$ errors on both the edge qubits or $\hat{Z}$ errors on penultimate qubits. This doesn't mean the $1^{\rm st}/N^{\rm th}$ qubit are immune to $\hat{X}$ errors, rather it is a by-product of how we are determining error probabilities. The $\hat{X}$ error on the first qubit here will appear as a slightly higher $\hat{Z}$ error rate on the 2nd qubit. Similarly in GHZ states, we have stabiliser generators $\hat{Z}_i\hat{Z}_{1} ~\textrm{for}~ i \in [2,N]$. We choose the first qubit to be the `central qubit' whose $\hat{Z}$ error we consider equivalent to a $\hat{Z} $ error on other 'boundary qubits'. We then consider $\hat{Z}$ errors to be only possible on the central qubit and not the other $N-1$ 'boundary qubits'. This makes the number of error probabilities to estimate $2N+1$. For a $N$-long branched chain, with $M$ branches on each qubit of the chain, we have  stabilisers that look like $\hat{X}_{\rm Branch} \hat{Z}_{\rm Central}$. Similar to GHZ states, we do not estimate the branch qubits $\hat{Z}$ errors. This results in the number of errors to be estimated as $2MN + 3N$. Since $\hat{Z}$ error on the central qubit can equivalently act on any of the other boundary qubits, dividing this quantity by $N$  for GHZ states (and equivalently $M+1$ for branched chains) gives us a per qubit $\hat{Z}$ error.

Calculating all $3N- 2$ error probabilities for small ($N<100$) chains, we find that the errors on bulk qubits are identical. This is also physically motivated since the pulse-cycle remains identical for each photon. For  larger chains we directly assume the errors on bulk qubits are identical to each other, and similarly errors on boundary qubits are identical to each other as well. This is a reasonable assumption in the limit of large chains approaching the thermodynamic limit, which have a larger bulk and lesser position/emission time-dependent variations, which are finite size effects. This is motivated by the fact we would like to ideally produce large resource states. This decreases the number of parameters to estimate, and hence stabiliser expectation values needed to estimate all errors. This means we now have 5 parameters to estimate for the entire state, three for qubits in the bulk and two for the 'boundary qubits'. Without making the assumption of identical or uncorrelated qubits, one can use methods like log-linear analysis \cite{Agresti2013JohnWiley} on the vector $\vec{p}$ of all $2^N$ probabilities calculated above, to obtain individual error probabilities on qubits and correlated errors.

Using the methods and assumptions described above, we have calculated the errors up to first order in experimentally determinable parameters for both three-level and four-level emitters. For three-level emitters, the Pauli errors are dependent on $|\gamma|^2$, $\bar{\eta} = 1-\eta$ and $\bar{\zeta} = 1-\zeta$ , where $|\gamma|^2$ is the average photon number of the excitation laser that leaks into our state, ideally 0. Here, $\eta$ is the efficiency of the emitter, ideally 1, and $\zeta$ is the visibility like term, which is also ideally 1. The Pauli error probabilities dependent on these parameters are listed in Table~\ref{tab:3_4_Level} . We obtain a biased error model where only the $Z$ errors are affected by $\zeta$, this is consistent with phonons causing dephasing errors on the qubit. All errors are affected by a $\bar{\eta}|\gamma|^2$ term, which is when the emitted single photon is lost, and replaced a photon from the coherent laser, which is also undetectable by other multi-photon heralded errors. This acts as uniform depolarising noise, i.e. affects all error channels equivalently. Note that $\bar{\eta}$ or $|\gamma|^2$ terms do not appear individually, as they take us out of the one photon subspace and can be heralded. We find that the errors are the same for $N$-long chains, branched chains, or $N$-qubit GHZ states in the limit of large $N$, where we distribute $\hat{Z}$ errors per qubit as described previously.

For four-level emitters, the Pauli errors are dependent on $|\gamma|^2$, $\bar{\eta} = 1-\eta$ and $\bar{\mathcal{D}} = 1-\mathcal{D}$, where $\bar{\mathcal{D}}$ is the birefringence of the emitter, which is ideally 0. The Pauli error probabilities dependent on these parameters are listed in Table~\ref{tab:3_4_Level} . Here we also have biased errors, where the $X$ errors are primarily affected by $\mathcal{D}$. We can understand this from observing the affect of $\mathcal{D}$ on $\ket{\bar{0}/\bar{1}}$, as seen in Eq.~\eqref{01barDef}.  As $\mathcal{D}$ deviates from 1, the $\ket{\bar{0}}~(\ket{\bar{1}})$ state has a greater overlap with $\ket{R}~(\ket{L})$, i.e. it acts like bit-flip noise. All errors here are also affected by a $\bar{\eta}|\gamma|^2$ term, but with the half the value as in three-level emitters. This is because two laser pulses are required to produce one qubit in three-level emitters, as opposed to a single pulse in four-level emitters. We also find this emitter shows odd-even effects for odd $N$, in branched chain and chain states. For odd $N$, the errors at even qubit positions are slightly more than that at odd positioned states. However the average error over odd and even positions is the same as that for even $N$, which we have reported in Table~\ref{tab:3_4_Level}. As for three-level emitters, we find the errors for $N$-qubit GHZ and $N$-long branched chains is the same as $N$-long chains.

\renewcommand{\arraystretch}{1.5}
\begin{table}[t]
    \centering
    \begin{tabular}{|c|c|c|}
        \hline
         &  Three-level emitter &Four-level emitter\\
         \hline
         $p_x$&   $\frac{\bar{\eta}|\gamma|^2}{2}$ &$\frac{\bar{\eta}|\gamma|^2}{4} + \bar{\mathcal{D}}$\\
         $p_y$&   $\frac{\bar{\eta}|\gamma|^2}{2}$ &$\frac{\bar{\eta}|\gamma|^2}{4}$\\
         $p_z$&   $\frac{\bar{\eta}|\gamma|^2}{2}+\frac{\bar{\zeta}}{2}$ &$\frac{\bar{\eta}|\gamma|^2}{4}$\\
         \hline
    \end{tabular}
    \caption{First order approximations of error probabilities per qubit for the different level emitters in Sec. \ref{subsec:Wavefuncions4} and Sec.\ref{subsec:Wavefuncions3}. Errors reported are the same for the three types of states considered; a $N$-long chain, an $N$-qubit GHZ state or an $N$-long branched-chain, each qubit with $M$ branches. For GHZ states and Branched chains, per qubit error is inferred by dividing the total Z error rate by $N$ or $M+1$ respectively. For Chain/Branched Chain of the four-level emitter we have only reported errors for even $N$.}
    \label{tab:3_4_Level}
\end{table}

\section{Fusion} \label{sec:Fusion}

In FBQC, resource states and two-qubit measurements are the fundamental components to perform computations. The two-qubit measurements here are referred to as type-two fusion~\cite{Browne2005PhysRevLett}. When performed with linear optical circuits type-two fusion is inherently a probabilistic measurement. When successfully measured it yields both the $\hat{X}\hat{X}$ and $\hat{Z}\hat{Z}$ Pauli measurement outcomes. The fusion measurement can also fail, in which case we obtain only one of the values, $\hat{X}\hat{X}$ or $\hat{Z}\hat{Z}$ and can be biased to either one \cite{Bombin2023arXiv}.

The combination of fusion measurements form stabiliser measurements, which subsequently combine to form $\hat{Z}$ or $\hat{X}$ checks \cite{Gottesman1997ArXiv}, the structure of which is informed by the choice of the error-correcting code. For dual-rail qubits in photonic circuits, these fusion measurements can be performed with linear optical elements and photon counting detectors. Success or failure outcomes are heralded by detection patterns that depend on the encoding of the state and the Bell state onto which it is projected. In this scheme we can successfully perform a fusion with a probability of a 1/2 \cite{Calsamiglia2001AppPhysB}. One such circuit is shown in Fig.\ref{fig:levelsOverall}, along with  one of the detection patterns that heralds the $\ket{\Phi^+}$ Bell state.

We now discuss how to model fusion measurements with error models so far. After a fusion measurement each detection pattern corresponds to a different noise map on the neighbouring photons. This map corresponds to an ideal Bell projection plus Pauli noise. As in state preparation, we assume the errors are Pauli-like on account of using error mitigation techniques like Pauli twirling \cite{Knill2008PhysRevA}, or randomised compiling \cite{Wallman2016PhysRevA}. The error on states after fusion is a function of the errors on the input states. Here we calculate the mapping for a Bell-state to get a sense of how fusion affects errors on states. For the general state, additional errors by fusion would explicitly depend on the structure on the input state.

We are going to do a simple entanglement swapping thought experiment. We input one photon from each of two $\ket{\Phi^+}$ Bell states into the fusion circuit shown in Fig.~\ref{fig:levelsOverall}(c), and detect the $\ket{1_x1_y00}$  (or $\ket{001_x1_y})$) state, i.e. photons are detected in the first (last) two rails, of unspecified temporal modes, and no photons are detected in rails three and four (one and two). The remaining photons of the original Bell states are then in a mixed state close to the $\ket{\Phi^+}$ Bell state. The wavefunctions of the input Bell states are known from Sec.~\ref{sec:Wavefuncions}, and propagated through the fusion circuit with lossless and balanced beamsplitters, and then projected on the detected state. Additionally we assume the detectors cannot differentiate between the coherent state and the emitted photon. The density matrix of the remaining photons is then calculated by tracing out the environment and loss channels. Since the ideal output is a $\ket{\Phi^+}$ Bell state, we calculate the relevant stabiliser expectation values, i.e. $\langle\hat{I}\hat{I}\rangle, \langle\hat{X}\hat{X}\rangle, \langle-\hat{Y}\hat{Y}\rangle ~\text{and}~ \langle\hat{Z}\hat{Z}\rangle$, as in Sec.~\ref{sec:StabExp}, and infer Pauli errors acting on the output state. Comparing the Pauli errors on the output state with those of the input state (known from Sec.~\ref{subsec:PauliErrFromStab}) we can calculate the applied Pauli map from the fusion, which is explicitly calculated in Appendix \ref{FusionMapping}. Just as in the case of state generation, there are certain Pauli errors we cannot distinguish from each other and thus have the freedom to associate the error with either of the output states. Since the Bell state is symmetric with respect to both qubits, we will choose the qubit that the error channels act on.

We expect the output of fusion measurements to have additional errors, despite perfect linear optical elements, since the input states to the fusion are imperfect, i.e. the states we've described in Sec.~\ref{sec:Wavefuncions}. By additional errors we refer to errors that will not be captured by a simple model of imperfect resource states with Pauli errors undergoing perfect fusion. For example, the visibility of the photons will affect the outcome of the fusion, since the photons will not interfere perfectly on a beam splitter. This is despite visibility errors not appearing as Pauli errors in states produced by the four-level emitter (see Table~\ref{tab:3_4_Level}). Furthermore, the leaked coherent photon can be detected, even if the emitted photon is lost, resulting in a false detection event. We therefore refer to such errors as added fusion errors.

Calculating the Pauli map for three- and four-level emitters we obtain the probability of added fusion errors $p_x$, $p_y$, and $p_z$, which are the probabilities of $\{\hat{X},\hat{Y},\hat{Z}\}$ respectively being applied to the noisy entangled state after the fusion measurement. We also define $p_o=1 - p_x - p_y -p_z$ as the probability of no added errors. The probability of fusion success is also reduced for these error models. Note that these error probabilities are in addition to Pauli errors that occur as a part of state preparation. The values approximated to first order are listed in Table \ref{tab:FusionMaps}  in terms of the loss $\bar{\eta} = 1-\eta$, coherent laser leakage $|\gamma|^2$  and distinguishability $\mathcal{\bar{V}} = 1-\mathcal{V}$, where $\mathcal{V} = \sum_i |\alpha_i|^4$ is the single-photon Hong-Ou-Mandel visibility. We have reported the mapping averaged over the two success outcomes, for $\ket{\Phi^+}$ and $\ket{\Phi^-}$ Bell states. We find the mapping for three-level emitters when approximated to first order, does not depend on the parameter $\zeta$, the visibility-like term. It does however show up in higher order terms. Here we see errors with the same depolarising term as before $\bar{\eta}|\gamma|^2$. In four-level emitters we also see an explicit dependence on visibility. For the individual success outcomes we see a dependence on birefringence $\bar{\mathcal{D}} = 1- \mathcal{D}$, but not when averaged with probability of detection (see Appendix \ref{ExtraFusion}).

We have also looked at the error maps for the detection patterns corresponding to fusion failure. The detected states are those with a two-photon detection in a rail, with none in the rest, e.g. $\ket{2_x 000}$, though it is not necessary both detected photons be of the same temporal-mode. Only $\hat{X}$ errors are detectable upon fusion failure, since the states produced are stabilised by $\hat{I}\hat{Z}$, $\hat{Z}\hat{I}$, $\hat{Z}\hat{Z}$ and $\hat{I}\hat{I}$.
For three-level emitters we only have $\hat{X}$ errors with probability $\bar{\eta}|\gamma|^2$ on each surviving photon. Similarly only $\hat{X}$ errors occur on fusion failure, which are dependent on $\mathcal{D}$ as well. This calculation and all error rates are explicitly listed in Appendix \ref{ExtraFusion}. 

\renewcommand{\arraystretch}{1.5}
\begin{table}[!t]
    \centering
    \begin{tabular}{|c|c|c|}
        \hline
         &  Three-Level& Four-Level\\
         \hline
         Prob. of Detection &  $\frac{1}{2}(1-2\bar{\eta}+8\bar{\eta}|\gamma|^2)$& $\frac{1}{2}(1-2\bar{\eta}+4\bar{\eta}\vert\gamma\vert^2)$\\
 $p_o$& $1 - 3\bar{\eta}|\gamma|^2$&$1  -\frac{3}{2}  \bar{\eta}|\gamma|^2-\frac{\bar{\mathcal{V}}}{2}$\\
 $p_x$& $\bar{\eta}|\gamma|^2$&$\frac{1}{2}  \bar{\eta}|\gamma|^2$\\
         $p_y$&  $\bar{\eta}|\gamma|^2$& $\frac{1}{2}\bar{\eta}|\gamma|^2$\\
         $p_z$&  $\bar{\eta}|\gamma|^2$& $\frac{1}{2}  \bar{\eta}|\gamma|^2+\frac{\bar{\mathcal{V}}}{2}$\\
         \hline
    \end{tabular}
    \caption{Probability of application of $\hat{I}, \hat{X}, \hat{Y} ~\text{and} ~\hat{Z}$ respectively for the Pauli map applied due to fusion measurements, for Bell states emitted from three- and four-level emitters.}
    \label{tab:FusionMaps}
\end{table}

\section{Discussion and Conclusion} \label{sec:discussionAndConclusion}

We have developed an  end-to-end  theoretical approach, connecting observables of single-photon sources to Pauli error rates relevant for QEC. We derived expressions for the complete wave function of any emitted state, represented compactly as MPSs and calculated exact stabiliser expectation values and error rates, as well as the mapping applied after fusion measurements. 

The error rates on emitted caterpillar states, are affected by a uniform depolarising noise due to coherent leakage, which is reduced by a factor of two in four-level emitters. This possibly represents a meaningful advantage in selecting four-level emitters for experimental implementations. Furthermore, we see the noise is also biased, depending on the type of emitter, $\hat{Z}$ errors in three-level emitters, and $\hat{X}$ errors in four-level emitters. For QEC purposes this is optimistic: thresholds under biased noise models are known to exceed those under purely depolarising noise for certain codes \cite{Sahay2023PhysRevLet}. Additionally the nature of errors points experimentalists to the critical aspects of emitters to optimize. Fidelity alone, as a scalar measure of state quality, cannot capture the asymmetry of errors, which is critical for code selection.

\renewcommand{\arraystretch}{1.5}
\begin{table}[!t]
    \centering
    \begin{tabular}{|c|c|c|}
        \hline
         &  Three-Level& Four-Level\\
         \hline
         $\eta$&  $>$ 0.967& $>$ 0.933\\
 $\mathcal{V}$& -&$>$ 0.996\\
         $\zeta$&  $>$ 0.998& -\\
         $\mathcal{D}$&  -& $>$ 0.999\\
         \hline
    \end{tabular}
    \caption{Parameter requirements for three- and four-level emitters to achieve $<0.1\%$ error per qubit for each error source. Here, $\eta$ is the total photonic efficiency, $\mathcal{V}$ is the Hong-Ou-Mandel visibility of the single photons produced, $\zeta$ is a visibility-like parameter for three-level emitter, and $1-\mathcal{D}$ is the birefringence parameter for the four-level system.}
    \label{tab:ParamLimits}
\end{table}

Assuming all produced states are resources for FBQC, and each photon is measured in a fusion measurement, the estimated total error on a single photon is approximately the sum of state and fusion-induced errors  we have calculated here. While the fusion errors calculated here are explicitly for Bell states, this is still informative to obtain order of magnitude estimates. We now use this to obtain limits on experimental parameters. The error thresholds for well-studied QEC algorithms, e.g. the surface code are $\approx 1\%$ \cite{Fowler2012PhysRevA, Bartolucci2021NatComm, Paesani2023PhysRevLet}. Taking the operating point to be an optimistic error rate of $0.5\%$, we assign an error budget of $<0.1\%$ to each source of error per qubit. This allows us to calculate lower bounds for the parameters. This is reasonable considering we have assumed perfect linear optical elements in the circuit, which will likely add errors as well in real circuits. State-of-the-art photon sources have a single-photon purity of $g^{(2)}(0)\sim 0.01$ \cite{Tomm2021Nature}. Here we assume that this results from laser leakage i.e. $|\gamma|^2 = 0.01$. For three-level emitters this gives us bounds for efficiency $\eta > 0.967$, and for the visibility-like parameter $\bar{\zeta} < 0.002$. For four-level emitters, the requirement on efficiency is more relaxed: $\eta > 0.933$.  however, the bounds on birefringence are very stringent $\bar{\mathcal{D}} < 9 \times 10^{-4}$ and visibility $\mathcal{V} > 0.996$. We have listed all these constraints compactly in Table~\ref{tab:ParamLimits}. We note that the efficiency is also constrained by the levels of fusion erasure that can be tolerated by an error-correcting code which we do not considered here \cite{Varnava2008PhysRevLet, Bartolucci2025arXiv}. These stringent experimental constraints on four-level emitters make them less attractive for the purposes of FBQC. Further we note that if depolarising noise can be traded for erasure \cite{Stace2009PhysRevLett, Bartolucci2021NatComm}, it would be beneficial in terms of error correction. This is especially for states produced by emitters which suffer from depolarising noise as we have shown. Definitive comparison between the types of emitters however will require more detailed analysis; this is highly dependent on the code implemented and whether any additional encoding is employed \cite{Bartolucci2025arXiv}.

While we have largely considered noise that influences photon emission, our models can be extended to integrate additional types of noise such as spin noise, and finite branching ratios which are relevant for most quantum-emitter platforms \cite{Tiurev2021PhysRevA}. Additionally, while we use $G^{(1)}$ to reconstruct the photon wavefunction, we note that eventually only the parameters $\zeta$, $\eta$, $\cal D$, $\cal V$, and $\gamma$ enter into the Pauli error rates. Any measurement set that extracts these parameters will suffice to successfully quantify the Pauli error rates given the assumptions we made on the types of noise that are present.

Building on the work here one could explicitly use the states produced as resource states in error correction algorithms, computing thresholds in terms of experimental parameters. This can inform both theoreticians and experimentalists on emitter design and resource-state selection to optimise quantum computing schemes and minimise logical noise. 

\bibliography{references}

@article{Lindner2009PhysRevA,
  title = {A Photonic Cluster State Machine Gun},
  author = {Lindner, Netanel H. and Rudolph, Terry},
  year = {2009},
  month = {Sep},
  journal = {Physical Review Letters},
  shortjournal = {Phys. Rev. Lett.},
  volume = {103},
  number = {11},
  eprint = {0810.2587},
  eprinttype = {arXiv},
  eprintclass = {quant-ph},
  pages = {113602},
  issn = {0031-9007, 1079-7114},
  doi = {10.1103/PhysRevLett.103.113602},
  url = {http://arxiv.org/abs/0810.2587},
  urldate = {2025-02-12}
}

@article{Tiurev2021PhysRevA,
  title = {Fidelity of time-bin-entangled multiphoton states from a quantum emitter},
  author = {Tiurev, Konstantin and Mirambell, Pol Llopart and Lauritzen, Mikkel Bloch and Appel, Martin Hayhurst and Tiranov, Alexey and Lodahl, Peter and S\o{}rensen, Anders S\o{}ndberg},
  journal = {Phys. Rev. A},
  volume = {104},
  issue = {5},
  pages = {052604},
  numpages = {21},
  year = {2021},
  month = {Nov},
  publisher = {American Physical Society},
  doi = {10.1103/PhysRevA.104.052604},
  url = {https://link.aps.org/doi/10.1103/PhysRevA.104.052604}
}

@article{Antoniadis2022NPJQuantInf,
   title={A chiral one-dimensional atom using a quantum dot in an open microcavity},
   volume={8},
   ISSN={2056-6387},
   url={http://dx.doi.org/10.1038/s41534-022-00545-z},
   DOI={10.1038/s41534-022-00545-z},
   number={1},
   journal={npj Quantum Information},
   publisher={Springer Science and Business Media LLC},
   author={Antoniadis, Nadia O. and Tomm, Natasha and Jakubczyk, Tomasz and Schott, Rüdiger and Valentin, Sascha R. and Wieck, Andreas D. and Ludwig, Arne and Warburton, Richard J. and Javadi, Alisa},
   year={2022},
   month=Mar }

@article{IlesSmith2017NPHOT,
   title={Phonon scattering inhibits simultaneous near-unity efficiency and indistinguishability in semiconductor single-photon sources},
   volume={11},
   ISSN={1749-4893},
   url={http://dx.doi.org/10.1038/nphoton.2017.101},
   DOI={10.1038/nphoton.2017.101},
   number={8},
   journal={Nature Photonics},
   publisher={Springer Science and Business Media LLC},
   author={Iles-Smith, Jake and McCutcheon, Dara P. S. and Nazir, Ahsan and Mørk, Jesper},
   year={2017},
   month=July, pages={521–526} }

@ARTICLE{Sollner2015NNANO,
  author = {Immo S\"{o}llner and Sahand Mahmoodian and Sofie Lindskov Hansen and Leonardo Midolo and Gabija Kirsanske and Tomamaso Pregnolato and Haitham El-Ella and Eun Hye Lee and Jin Dong Song and S{\o}ren Stobbe and Peter Lodahl},
  title = {Deterministic photon--emitter coupling in chiral photonic circuits},
  journal = {Nat. Nanotechnol.},
  volume = {10},
  pages = {775},
  year = {2015},
  doi = {10.1038/nnano.2015.159}
}

@article{Lodahl2017Nature,
  title={Chiral quantum optics},
  author={Lodahl, Peter and Mahmoodian, Sahand and Stobbe, S{\o}ren and Rauschenbeutel, Arno and Schneeweiss, Philipp and Volz, J{\"u}rgen and Pichler, Hannes and Zoller, Peter},
  journal={Nature},
  volume={541},
  number={7638},
  pages={473--480},
  year={2017},
  doi={doi:10.1038/nature21037}
}

@article{Lobl2025AmPhysSoc,
  title = {Generating graph states with a single quantum emitter and the minimum number of fusions},
  author = {L\"obl, Matthias C. and Pettersson, Love A. and Jena, Andrew and Dellantonio, Luca and Paesani, Stefano and S\o{}rensen, Anders S.},
  journal = {Phys. Rev. A},
  volume = {111},
  issue = {5},
  pages = {052604},
  numpages = {16},
  year = {2025},
  month = {May},
  publisher = {American Physical Society},
  doi = {10.1103/PhysRevA.111.052604},
  url = {https://link.aps.org/doi/10.1103/PhysRevA.111.052604}
}

@misc{Bartolucci2021NatComm,
  title = {Fusion-Based Quantum Computation},
  author = {Bartolucci, Sara and Birchall, Patrick and Bombin, Hector and Cable, Hugo and Dawson, Chris and {Gimeno-Segovia}, Mercedes and Johnston, Eric and Kieling, Konrad and Nickerson, Naomi and Pant, Mihir and Pastawski, Fernando and Rudolph, Terry and Sparrow, Chris},
  year = 2021,
  month = jan,
  number = {arXiv:2101.09310},
  eprint = {2101.09310},
  primaryclass = {quant-ph},
  doi = {10.48550/arXiv.2101.09310},
  urldate = {2023-07-14},
  archiveprefix = {arXiv},
}

@misc{Bartolucci2021Arxiv,
      title={Creation of Entangled Photonic States Using Linear Optics}, 
      author={Sara Bartolucci and Patrick M. Birchall and Mercedes Gimeno-Segovia and Eric Johnston and Konrad Kieling and Mihir Pant and Terry Rudolph and Jake Smith and Chris Sparrow and Mihai D. Vidrighin},
      year={2021},
      eprint={2106.13825},
      archivePrefix={arXiv},
      primaryClass={quant-ph},
      url={https://arxiv.org/abs/2106.13825}, 
}

@article{Raussendorf2003PhysRevA,
   title={Measurement-based quantum computation on cluster states},
   volume={68},
   ISSN={1094-1622},
   url={http://dx.doi.org/10.1103/PhysRevA.68.022312},
   DOI={10.1103/physreva.68.022312},
   number={2},
   journal={Physical Review A},
   publisher={American Physical Society (APS)},
   author={Raussendorf, Robert and Browne, Daniel E. and Briegel, Hans J.},
   year={2003},
   month=aug }

@article{Paesani2023PhysRevLet,
  title = {High-Threshold Quantum Computing by Fusing One-Dimensional Cluster States},
  author = {Paesani, Stefano and Brown, Benjamin J.},
  journal = {Phys. Rev. Lett.},
  volume = {131},
  issue = {12},
  pages = {120603},
  numpages = {7},
  year = {2023},
  month = {Sep},
  publisher = {American Physical Society},
  doi = {10.1103/PhysRevLett.131.120603},
  url = {https://link.aps.org/doi/10.1103/PhysRevLett.131.120603}
}

@article{Zhang2019Light,
  title = {Generation of Multiphoton Quantum States on Silicon},
  author = {Zhang, Ming and Feng, Lan-Tian and Zhou, Zhi-Yuan and Chen, Yang and Wu, Hao and Li, Ming and Gao, Shi-Ming and Guo, Guo-Ping and Guo, Guang-Can and Dai, Dao-Xin and Ren, Xi-Feng},
  year = 2019,
  month = may,
  journal = {Light: Science \& Applications},
  volume = {8},
  number = {1},
  pages = {41},
  issn = {2047-7538},
  doi = {10.1038/s41377-019-0153-y},
}

@article{Schwartz2016Scien,
author = {I. Schwartz  and D. Cogan  and E. R. Schmidgall  and Y. Don  and L. Gantz  and O. Kenneth  and N. H. Lindner  and D. Gershoni },
title = {Deterministic generation of a cluster state of entangled photons},
journal = {Science},
volume = {354},
number = {6311},
pages = {434-437},
year = {2016},
doi = {10.1126/science.aah4758},
URL = {https://www.science.org/doi/abs/10.1126/science.aah4758},
eprint = {https://www.science.org/doi/pdf/10.1126/science.aah4758}}

@article{Kuhn2002AmPhysSoc,
  title = {Deterministic Single-Photon Source for Distributed Quantum Networking},
  author = {Kuhn, Axel and Hennrich, Markus and Rempe, Gerhard},
  journal = {Phys. Rev. Lett.},
  volume = {89},
  issue = {6},
  pages = {067901},
  numpages = {4},
  year = {2002},
  month = {Jul},
  publisher = {American Physical Society},
  doi = {10.1103/PhysRevLett.89.067901},
  url = {https://link.aps.org/doi/10.1103/PhysRevLett.89.067901}
}

@article{Cogan2023NatPhot,
  title = {Deterministic Generation of Indistinguishable Photons in a Cluster State},
  author = {Cogan, Dan and Su, Zu-En and Kenneth, Oded and Gershoni, David},
  year = {2023},
  month = {Apr},
  journal = {Nature Photonics},
  shortjournal = {Nature Photonics},
  volume = {17},
  number = {4},
  pages = {324--329},
  issn = {1749-4893},
  doi = {10.1038/s41566-022-01152-2},
  url = {https://doi.org/10.1038/s41566-022-01152-2},
}

@article{Li2022NPJQuantInf,
   title={Photonic resource state generation from a minimal number of quantum emitters},
   volume={8},
   ISSN={2056-6387},
   url={http://dx.doi.org/10.1038/s41534-022-00522-6},
   DOI={10.1038/s41534-022-00522-6},
   number={1},
   journal={npj Quantum Information},
   publisher={Springer Science and Business Media LLC},
   author={Li, Bikun and Economou, Sophia E. and Barnes, Edwin},
   year={2022},
   month=feb }

@article{Panos2008PRA,
  title = {Fault-tolerant quantum computation against biased noise},
  author = {Aliferis, Panos and Preskill, John},
  journal = {Phys. Rev. A},
  volume = {78},
  issue = {5},
  pages = {052331},
  numpages = {9},
  year = {2008},
  month = {Nov},
  publisher = {American Physical Society},
  doi = {10.1103/PhysRevA.78.052331},
  url = {https://link.aps.org/doi/10.1103/PhysRevA.78.052331}
}

@article{Tuckett2019PRX,
  title = {Tailoring Surface Codes for Highly Biased Noise},
  author = {Tuckett, David K. and Darmawan, Andrew S. and Chubb, Christopher T. and Bravyi, Sergey and Bartlett, Stephen D. and Flammia, Steven T.},
  journal = {Phys. Rev. X},
  volume = {9},
  issue = {4},
  pages = {041031},
  numpages = {22},
  year = {2019},
  month = {Nov},
  publisher = {American Physical Society},
  doi = {10.1103/PhysRevX.9.041031},
  url = {https://link.aps.org/doi/10.1103/PhysRevX.9.041031}
}

@article{Kiraz2004PhysRevA,
  title = {Quantum-Dot Single-Photon Sources: {{Prospects}} for Applications in Linear Optics Quantum-Information Processing},
  shorttitle = {Quantum-Dot Single-Photon Sources},
  author = {Kiraz, A. and Atatüre, M. and Imamoğlu, A.},
  year = {2004},
  month = {Mar},
  journal = {Physical Review A},
  shortjournal = {Phys. Rev. A},
  volume = {69},
  number = {3},
  pages = {032305},
  issn = {1050-2947, 1094-1622},
  doi = {10.1103/PhysRevA.69.032305},
  url = {https://link.aps.org/doi/10.1103/PhysRevA.69.032305},
  urldate = {2025-02-20},
  langid = {english},
}

@article{Wang2025NCOMM,
   title={Coherence in resonance fluorescence},
   volume={16},
   ISSN={2041-1723},
   url={http://dx.doi.org/10.1038/s41467-025-61884-x},
   DOI={10.1038/s41467-025-61884-x},
   number={1},
   journal={Nature Communications},
   publisher={Springer Science and Business Media LLC},
   author={Wang, Xu-Jie and Huang, Guoqi and Li, Ming-Yang and Wang, Yuan-Zhuo and Liu, Li and Wu, Bang and Liu, Hanqing and Ni, Haiqiao and Niu, Zhichuan and Ji, Weijie and Jiao, Rongzhen and Yin, Hua-Lei and Yuan, Zhiliang},
   year={2025},
   month=July }

@phdthesis{Sparrow2018ICL,
  title={Quantum interference in universal linear optical devices for quantum computation and simulation},
  author={Sparrow, Christopher},
  year={2018},
  school={Imperial College London}
}

@article{Thomas2022Nature,
   title={Efficient generation of entangled multiphoton graph states from a single atom},
   volume={608},
   ISSN={1476-4687},
   url={http://dx.doi.org/10.1038/s41586-022-04987-5},
   DOI={10.1038/s41586-022-04987-5},
   number={7924},
   journal={Nature},
   publisher={Springer Science and Business Media LLC},
   author={Thomas, Philip and Ruscio, Leonardo and Morin, Olivier and Rempe, Gerhard},
   year={2022},
   month=Aug, pages={677–681} }

@article{Qin2015Nature,
  title = {Complete Temporal Characterization of a Single Photon},
  author = {Qin, Zhongzhong and Prasad, Adarsh S. and Brannan, Travis and MacRae, Andrew and Lezama, A. and Lvovsky, A. I.},
  year = {2015},
  month = {Jun},
  journal = {Light: Science \& Applications},
  shortjournal = {Light Sci Appl},
  volume = {4},
  number = {6},
  pages = {e298-e298},
  publisher = {Nature Publishing Group},
  issn = {2047-7538},
  doi = {10.1038/lsa.2015.71},
  url = {https://www.nature.com/articles/lsa201571},
  urldate = {2024-10-29},
  langid = {english},
}

@article{Calsamiglia2001AppPhysB,
  title = {Maximum Efficiency of a Linear-Optical {{Bell-state}} Analyzer},
  author = {Calsamiglia, John and L{\"u}tkenhaus, Norbert},
  year = 2001,
  month = jan,
  journal = {Applied Physics B},
  volume = {72},
  number = {1},
  eprint = {quant-ph/0007058},
  pages = {67--71},
  issn = {0946-2171, 1432-0649},
  doi = {10.1007/s003400000484},
  urldate = {2024-12-31},
  archiveprefix = {arXiv}
}

@article{Tomm2021Nature,
  title = {A Bright and Fast Source of Coherent Single Photons},
  author = {Tomm, Natasha and Javadi, Alisa and Antoniadis, Nadia Olympia and Najer, Daniel and Löbl, Matthias Christian and Korsch, Alexander Rolf and Schott, Rüdiger and Valentin, Sascha René and Wieck, Andreas Dirk and Ludwig, Arne and Warburton, Richard John},
  year = {2021},
  month = {Apr},
  journal = {Nature Nanotechnology},
  shortjournal = {Nat. Nanotechnol.},
  volume = {16},
  number = {4},
  pages = {399--403},
  publisher = {Nature Publishing Group},
  issn = {1748-3395},
  doi = {10.1038/s41565-020-00831-x},
  url = {https://www.nature.com/articles/s41565-020-00831-x},
  urldate = {2025-02-20},
  langid = {english},
}

@article{Alexander2025Nature,
  title = {A Manufacturable Platform for Photonic Quantum Computing},
  author = {Alexander, Koen and Benyamini, Avishai and Black, Dylan and Bonneau, Damien and Burgos, Stanley and Burridge, Ben and Cable, Hugo and Campbell, Geoff and Catalano, Gabriel and Ceballos, Alejandro and Chang, Chia-Ming and Choudhury, Sourav Sen and Chung, C. J. and Danesh, Fariba and Dauer, Tom and Davis, Michael and Dudley, Eric and {Er-Xuan}, Ping and Fargas, Josep and Farsi, Alessandro and Fenrich, Colleen and Frazer, Jonathan and Fukami, Masaya and Ganesan, Yogeeswaran and Gibson, Gary and {Gimeno-Segovia}, Mercedes and Goeldi, Sebastian and Goley, Patrick and Haislmaier, Ryan and Halimi, Sami and Hansen, Paul and Hardy, Sam and Horng, Jason and House, Matthew and Hu, Hong and Jadidi, Mehdi and Jain, Vijay and Johansson, Henrik and Jones, Thomas and Kamineni, Vimal and Kelez, Nicholas and Koustuban, Ravi and Kovall, George and Krogen, Peter and Kumar, Nikhil and Liang, Yong and LiCausi, Nicholas and Llewellyn, Dan and Lokovic, Kimberly and Lovelady, Michael and Manfrinato, Vitor Riseti and Melnichuk, Ann and Mendoza, Gabriel and Moores, Brad and Mukherjee, Shaunak and Munns, Joseph and Musalem, Francois-Xavier and Najafi, Faraz and O'Brien, Jeremy L. and Ortmann, J. Elliott and Pai, Sunil and Park, Bryan and Peng, Hsuan-Tung and Penthorn, Nicholas and Peterson, Brennan and Peterson, Gabriel and Poush, Matt and Pryde, Geoff J. and Ramprasad, Tarun and Ray, Gareth and Rodriguez, Angelita Viejo and Roxworthy, Brian and Rudolph, Terry and Saunders, Dylan J. and Shadbolt, Pete and Shah, Deesha and Bahgat Shehata, Andrea and Shin, Hyungki and Sinsky, Jeffrey and Smith, Jake and Sohn, Ben and Sohn, Young-Ik and Son, Gyeongho and Souza, Mario C. M. M. and Sparrow, Chris and Staffaroni, Matteo and Stavrakas, Camille and Sukumaran, Vijay and Tamborini, Davide and Thompson, Mark G. and Tran, Khanh and Triplett, Mark and Tung, Maryann and Veitia, Andrzej and Vert, Alexey and Vidrighin, Mihai D. and Vorobeichik, Ilya and Weigel, Peter and Wingert, Matthew and Wooding, Jamie and Zhou, Xinran and {PsiQuantum team}},
  year = 2025,
  month = May,
  journal = {Nature},
  volume = {641},
  number = {8064},
  pages = {876--883},
  publisher = {Nature Publishing Group},
  issn = {1476-4687},
  doi = {10.1038/s41586-025-08820-7},
  urldate = {2026-05-24},
  copyright = {2025 The Author(s)},
}

@article{Sipahigil2014PhysRevLet,
  title = {Indistinguishable Photons from Separated Silicon-Vacancy Centers in Diamond},
  author = {Sipahigil, A. and Jahnke, K. D. and Rogers, L. J. and Teraji, T. and Isoya, J. and Zibrov, A. S. and Jelezko, F. and Lukin, M. D.},
  journal = {Phys. Rev. Lett.},
  volume = {113},
  issue = {11},
  pages = {113602},
  numpages = {5},
  year = {2014},
  month = {Sep},
  publisher = {American Physical Society},
  doi = {10.1103/PhysRevLett.113.113602},
  url = {https://link.aps.org/doi/10.1103/PhysRevLett.113.113602}
}

@article{Wilk2007Science,
author = {Tatjana Wilk  and Simon C. Webster  and Axel Kuhn  and Gerhard Rempe },
title = {Single-Atom Single-Photon Quantum Interface},
journal = {Science},
volume = {317},
number = {5837},
pages = {488-490},
year = {2007},
URL = {https://www.science.org/doi/abs/10.1126/science.1143835}}

@book{Agresti2013JohnWiley,
  title={Categorical Data Analysis},
  author={Agresti, Alan},
  edition={3rd},
  year={2013},
  publisher={John Wiley \& Sons},
  address={Hoboken, New Jersey},
  isbn={978-0-470-46363-5}
}

@phdthesis{Gottesman1997ArXiv,
      title={Stabilizer Codes and Quantum Error Correction}, 
      author={Daniel Gottesman},
      school = {California Institute of Technology},
      year={1997},
      eprint={quant-ph/9705052},
      archivePrefix={arXiv},
      primaryClass={quant-ph},
      url={https://arxiv.org/abs/quant-ph/9705052}, 
}

@article{Varnava2008PhysRevLet,
  title = {How Good Must Single Photon Sources and Detectors Be for Efficient Linear Optical Quantum Computation?},
  author = {Varnava, Michael and Browne, Daniel E. and Rudolph, Terry},
  journal = {Phys. Rev. Lett.},
  volume = {100},
  issue = {6},
  pages = {060502},
  numpages = {4},
  year = {2008},
  month = {Feb},
  publisher = {American Physical Society},
  doi = {10.1103/PhysRevLett.100.060502},
  url = {https://link.aps.org/doi/10.1103/PhysRevLett.100.060502}
}

@article{Schollwock2011Annals,
   title={The density-matrix renormalization group in the age of matrix product states},
   volume={326},
   ISSN={0003-4916},
   url={http://dx.doi.org/10.1016/j.aop.2010.09.012},
   DOI={10.1016/j.aop.2010.09.012},
   number={1},
   journal={Annals of Physics},
   publisher={Elsevier BV},
   author={Schollwöck, Ulrich},
   year={2011},
   month=Jan, pages={96–192} 
   }

@article{Sahay2023PhysRevLet,
  title = {Tailoring Fusion-Based Error Correction for High Thresholds to Biased Fusion Failures},
  author = {Sahay, Kaavya and Claes, Jahan and Puri, Shruti},
  journal = {Phys. Rev. Lett.},
  volume = {131},
  issue = {12},
  pages = {120604},
  numpages = {5},
  year = {2023},
  month = {Sep},
  publisher = {American Physical Society},
  doi = {10.1103/PhysRevLett.131.120604},
  url = {https://link.aps.org/doi/10.1103/PhysRevLett.131.120604}
}

@misc{Bartolucci2025arXiv,
      title={Comparison of schemes for highly loss tolerant photonic fusion based quantum computing}, 
      author={Sara Bartolucci and Tom Bell and Hector Bombin and Patrick Birchall and Jacob Bulmer and Christopher Dawson and Terry Farrelly and Samuel Gartenstein and Mercedes Gimeno-Segovia and Daniel Litinski and Yehua Liu and Robert Knegjens and Naomi Nickerson and Andrea Olivo and Mihir Pant and Ashlesha Patil and Sam Roberts and Terry Rudolph and Chris Sparrow and David Tuckett and Andrzej Veitia},
      year={2025},
      eprint={2506.11975},
      archivePrefix={arXiv},
      primaryClass={quant-ph},
      url={https://arxiv.org/abs/2506.11975}, 
}

@article{Knill2008PhysRevA,
   title={Randomized benchmarking of quantum gates},
   volume={77},
   ISSN={1094-1622},
   url={http://dx.doi.org/10.1103/PhysRevA.77.012307},
   DOI={10.1103/physreva.77.012307},
   number={1},
   journal={Physical Review A},
   publisher={American Physical Society (APS)},
   author={Knill, E. and Leibfried, D. and Reichle, R. and Britton, J. and Blakestad, R. B. and Jost, J. D. and Langer, C. and Ozeri, R. and Seidelin, S. and Wineland, D. J.},
   year={2008},
   month=Jan }

@article{Hauser2026NatComm,
  title = {Deterministic and highly indistinguishable single photons in the telecom C-band},
  volume = {17},
  ISSN = {2041-1723},
  url = {http://dx.doi.org/10.1038/s41467-026-68336-0},
  DOI = {10.1038/s41467-026-68336-0},
  number = {1},
  journal = {Nature Communications},
  publisher = {Springer Science and Business Media LLC},
  author = {Hauser,  Nico and Bayerbach,  Matthias and Kaupp,  Jochen and Reum,  Yorick and Peniakov,  Giora and Michl,  Johannes and Kamp,  Martin and Huber-Loyola,  Tobias and Pfenning,  Andreas T. and H\"{o}fling,  Sven and Barz,  Stefanie},
  year = {2026},
  month = Jan 
}

@book{West2001PrenticeHall,
  author    = {West, Douglas B.},
  title     = {Introduction to Graph Theory},
  edition   = {2nd},
  publisher = {Prentice Hall},
  address   = {Upper Saddle River, NJ},
  year      = {2001},
  isbn      = {978-0-13-014400-3}
}

@article{Wallman2016PhysRevA,
  title = {Noise tailoring for scalable quantum computation via randomized compiling},
  author = {Wallman, Joel J. and Emerson, Joseph},
  journal = {Phys. Rev. A},
  volume = {94},
  issue = {5},
  pages = {052325},
  numpages = {9},
  year = {2016},
  month = {Nov},
  publisher = {American Physical Society},
  doi = {10.1103/PhysRevA.94.052325},
  url = {https://link.aps.org/doi/10.1103/PhysRevA.94.052325}
}

@article{Browne2005PhysRevLett,
  title = {Resource-Efficient Linear Optical Quantum Computation},
  author = {Browne, Daniel E. and Rudolph, Terry},
  journal = {Phys. Rev. Lett.},
  volume = {95},
  issue = {1},
  pages = {010501},
  numpages = {4},
  year = {2005},
  month = {Jun},
  publisher = {American Physical Society},
  doi = {10.1103/PhysRevLett.95.010501},
  url = {https://link.aps.org/doi/10.1103/PhysRevLett.95.010501}
}

@article{Hein2004PhysRevA,
  title = {Multiparty entanglement in graph states},
  author = {Hein, M. and Eisert, J. and Briegel, H. J.},
  journal = {Phys. Rev. A},
  volume = {69},
  issue = {6},
  pages = {062311},
  numpages = {20},
  year = {2004},
  month = {Jun},
  publisher = {American Physical Society},
  doi = {10.1103/PhysRevA.69.062311},
  url = {https://link.aps.org/doi/10.1103/PhysRevA.69.062311}
}

@article{Fowler2012PhysRevA,
  title = {Surface codes: Towards practical large-scale quantum computation},
  author = {Fowler, Austin G. and Mariantoni, Matteo and Martinis, John M. and Cleland, Andrew N.},
  journal = {Phys. Rev. A},
  volume = {86},
  issue = {3},
  pages = {032324},
  numpages = {48},
  year = {2012},
  month = {Sep},
  publisher = {American Physical Society},
  doi = {10.1103/PhysRevA.86.032324},
  url = {https://link.aps.org/doi/10.1103/PhysRevA.86.032324}
}

@article{Buterakos2017PhysRevX,
  title = {Deterministic Generation of All-Photonic Quantum Repeaters from Solid-State Emitters},
  author = {Buterakos, Donovan and Barnes, Edwin and Economou, Sophia E.},
  journal = {Phys. Rev. X},
  volume = {7},
  issue = {4},
  pages = {041023},
  numpages = {10},
  year = {2017},
  month = {Oct},
  publisher = {American Physical Society},
  doi = {10.1103/PhysRevX.7.041023},
  url = {https://link.aps.org/doi/10.1103/PhysRevX.7.041023}
}

@article{Raussendorf2001PhysRevLett,
  title = {A One-Way Quantum Computer},
  author = {Raussendorf, Robert and Briegel, Hans J.},
  journal = {Phys. Rev. Lett.},
  volume = {86},
  issue = {22},
  pages = {5188--5191},
  numpages = {0},
  year = {2001},
  month = {May},
  publisher = {American Physical Society},
  doi = {10.1103/PhysRevLett.86.5188},
  url = {https://link.aps.org/doi/10.1103/PhysRevLett.86.5188}
}

@misc{Bombin2023arXiv,
      title={Increasing error tolerance in quantum computers with dynamic bias arrangement}, 
      author={Hector Bombín and Chris Dawson and Naomi Nickerson and Mihir Pant and Jordan Sullivan},
      year={2023},
      eprint={2303.16122},
      archivePrefix={arXiv},
      primaryClass={quant-ph},
      url={https://arxiv.org/abs/2303.16122}, 
}

@article{Lee2019IOPPubl,
  doi       = {10.1088/2058-9565/ab0a9b},
  url       = {https://doi.org/10.1088/2058-9565/ab0a9b},
  year      = {2019},
  month     = {mar},
  publisher = {IOP Publishing},
  volume    = {4},
  number    = {2},
  pages     = {025011},
  author    = {Lee, J P and Villa, B and Bennett, A J and Stevenson, R M and Ellis, D J P and Farrer, I and Ritchie, D A and Shields, A J},
  title     = {A quantum dot as a source of time-bin entangled multi-photon states},
  journal   = {Quantum Science and Technology},
}

@article{Hong1987PhysRevLett,
  title = {Measurement of subpicosecond time intervals between two photons by interference},
  author = {Hong, C. K. and Ou, Z. Y. and Mandel, L.},
  journal = {Phys. Rev. Lett.},
  volume = {59},
  issue = {18},
  pages = {2044--2046},
  numpages = {0},
  year = {1987},
  month = {Nov},
  publisher = {American Physical Society},
  doi = {10.1103/PhysRevLett.59.2044},
  url = {https://link.aps.org/doi/10.1103/PhysRevLett.59.2044}
}

@article{Zhai2022NatureNano,
   title={Quantum interference of identical photons from remote GaAs quantum dots},
   volume={17},
   ISSN={1748-3395},
   url={http://dx.doi.org/10.1038/s41565-022-01131-2},
   DOI={10.1038/s41565-022-01131-2},
   number={8},
   journal={Nature Nanotechnology},
   publisher={Springer Science and Business Media LLC},
   author={Zhai, Liang and Nguyen, Giang N. and Spinnler, Clemens and Ritzmann, Julian and Löbl, Matthias C. and Wieck, Andreas D. and Ludwig, Arne and Javadi, Alisa and Warburton, Richard J.},
   year={2022},
   month=May, pages={829–833} }

@misc{Xing2023arXiv,
      title={High-efficiency single-photon source above the loss-tolerant threshold for efficient linear optical quantum computing}, 
      author={Xing Ding and Yong-Peng Guo and Mo-Chi Xu and Run-Ze Liu and Geng-Yan Zou and Jun-Yi Zhao and Zhen-Xuan Ge and Qi-Hang Zhang and Hua-Liang Liu and Lin-Jun Wang and Ming-Cheng Chen and Hui Wang and Yu-Ming He and Yong-Heng Huo and Chao-Yang Lu and Jian-Wei Pan},
      year={2023},
      eprint={2311.08347},
      archivePrefix={arXiv},
      primaryClass={quant-ph},
      url={https://arxiv.org/abs/2311.08347}, 
}

@article{Stace2009PhysRevLett,
  title = {Thresholds for Topological Codes in the Presence of Loss},
  author = {Stace, Thomas M. and Barrett, Sean D. and Doherty, Andrew C.},
  journal = {Phys. Rev. Lett.},
  volume = {102},
  issue = {20},
  pages = {200501},
  numpages = {4},
  year = {2009},
  month = {May},
  publisher = {American Physical Society},
  doi = {10.1103/PhysRevLett.102.200501},
  url = {https://link.aps.org/doi/10.1103/PhysRevLett.102.200501}
}

\clearpage
\appendix
\onecolumngrid
\section{Calculating Fusion Mapping}\label{FusionMapping}

In this section we provide more details on the Pauli error map that is applied on the state after fusion and explicitly walk through the calculation of the density matrix after the Bell measurement.
As described in Sec. \ref{sec:Fusion} the input to the fusion measurement, as shown in Fig.\ref{fig:levelsOverall} ,  is two photons from two different emitted $\ket{\Phi^+}$ Bell states. With some probability we obtain a resultant state (here ideally $\ket{\Phi^+}$ ) of the remaining photons, for a detected pattern in the circuit ($\ket{1100}$ or $\ket{0011}$). Other detection patterns ($\ket{1010}$ / $\ket{0101}$ for success, and  $\ket{2000}$/$\ket{0002}$,$\ket{0200}$/$\ket{0020}$ for failure) indicate other resultant Bell states ($\ket{\Phi^-}$), or the occurrence of 'fusion failure', where we just obtain two disentangled photons measured out in the $\hat{X}/\hat{Z}$ basis. As we expanded upon in Sec.~\ref{sec:Fusion},  we model errors as the application of a Pauli map after an ideal fusion on two input states. Comparing error rates on the input states and output states (using the stabiliser expectation values), we can calculate the Pauli map applied and thus the added Pauli error due to fusion.

We illustrate the calculation for the four-level emitter, the same logic follows for three-level. The input is the emitted $\ket{\Phi^+}$ state expressed in $\ket{\bar{0}/\bar{1}}$ states, with some normalisation constant $\mathcal{N}$:
\begin{align*}
\ket{\Phi^+} &=  \frac{\ket{\bar{0}\bar{0}} + \ket{\bar{1}\bar{1}}}{\sqrt{2}\mathcal{N}} \\
&=  \frac{1}{\sqrt{2(1+k^2)}}  \ket{LL} + \ket{RR} + k\cdot (\ket{LR}+\ket{RL}) .\\
\end{align*}
Here we define, $k = 2\sqrt{\mathcal{D}(1-\mathcal{D})}$ , and obtain $\mathcal{N}^2 = 1+k^2 = 1 + 4{\mathcal{D}(1-\mathcal{D})}$. The complete input to the fusion operation, of two bell pairs is then the state :
\begin{align*}
\ket{\Phi^+}_{1,2} \otimes \ket{\Phi^+}_{3,4} = ~~ \frac{1}{2 \cdot(1+k^2)} ~~   &\ket{LL}_{23} ~ \left( \ket{LL} + k\ket{RL} + k\ket{LR} +k^2\ket{RR} \right)_{14} \\
+&\ket{RR}_{23} ~ \left( \ket{RR} + k\ket{RL} + k\ket{LR} +k^2\ket{LL} \right)_{14} \\
+&\ket{LR}_{23} ~ \left( \ket{LR} + k\ket{LL} + k\ket{RR} +k^2\ket{RL} \right)_{14} \\
+&\ket{RL}_{23} ~ \left( \ket{RL} + k\ket{LL} + k\ket{RR} +k^2\ket{LR} \right)_{14} .
\end{align*}
Here we've rearranged the photons to separate out the detected photons (denoted by subscript 2,3) and the remaining photons that do not pass through the circuit (denoted by subscript 1,4).
For the particular fusion circuit in Fig.~\ref{fig:levelsOverall}(c) , the detection patterns $\ket{1100}$ and $\ket{0011}$ at the detectors, successfully project the photons 1,4 onto state $\ket{\Phi^+}$ (ideally), which occurs with a probability 1/8 each.  We shall consider the detection pattern $\ket{1100}$, but the calculation remains identical for $\ket{0011}$. 
Let us say we detect a photon in the first two rails, in the modes $x,y$ respectively i.e. we detect  $\ket{1_x1_y00}$. We sum over the modes later, since the detectors cannot discern between different temporal modes. 
Rather than propagating the state through the circuit, we equivalently back-propagate the detected state to the front of the linear optical circuit and apply this on the input. This implies that the detected state, denoted as as $\ket{\rm Detect}_{x,y}=\ket{1_x1_y00}$ then projects on to photons 2,3 of the input state as,
\begin{align}
\ket{\rm Detect}_{x,y} = \frac{1}{2} ~~ \ket{1_x1_y00} + \ket{01_y01_x}
+ \ket{1_x01_y0} + \ket{001_y1_x}. 
\end{align}
We then have the unnormalised state,
\begin{align}
\hat{\rho}_{\rm out} = \sum_{x,y}\operatorname{Tr}_{\rm loss/environ}\left[~ \bra{\rm Detect}_{x,y}\left(\ket{\Phi^+}_{1,2} \otimes \ket{\Phi^+}_{3,4}\right)\left(\bra{\Phi^+}_{1,2} \otimes \bra{\Phi^+}_{3,4}\right) \ket{\rm Detect}_{x,y}\right].
\end{align}

After normalising by $\mathcal{T}$, we can write out the complete output density matrix as,
\begin{align}
\hat{\rho}_{\rm out} = \frac{1}{\mathcal{T} . 4(1+k^2)^2} ~~~~ 
   &(\ket{LL}\bra{LL}+ \ket{RR}\bra{RR}) ~~ ( \mathcal{A}  + \mathcal{A}k^4 + 2\mathcal{B}k^2 + 2\mathcal{C}k^2) \nonumber\\
+ ~& (\ket{LL}\bra{RR}+ h.c.) ~~ (2\mathcal{A}k^2 + \mathcal{B} + \mathcal{B}k^4 + 2\mathcal{C}k^2) \nonumber\\
+~ & (\ket{LL}\bra{LR} + \ket{LL}\bra{RL} +\ket{RR}\bra{LR} + \ket{RR}\bra{RL} + h.c.) ~~ (\mathcal{A} + \mathcal{B} + \mathcal{C})(k + k^3)\nonumber \\
+ ~& (\ket{LR}\bra{LR}+\ket{RL}\bra{RL}) ~~ (2\mathcal{A}k^2 + 2\mathcal{B}k^2 + \mathcal{C} + \mathcal{C}k^4)\nonumber \\
+ ~& \left(\ket{LR}\bra{RL}+h.c.\right) ~~ (2\mathcal{A}k^2 + 2\mathcal{B}k^2 + 2\mathcal{C}k^2 ).
\end{align}
Here we have used the following shorthand,
\begin{align}
    \mathcal{A} = ~&\frac{1}{8}\left(3|\gamma|^4(1-\eta)^2 + 4\eta|\gamma|^2(1-\eta) + 2\eta^2\right), \\
    \mathcal{B} = ~&\mathcal{V}\eta^2  \frac{1}{4}, \\
    \mathcal{C} = ~&\frac{1}{8}\left(3|\gamma|^2(1-\eta) + 4\eta\right).(|\gamma|^2(1-\eta)).
\end{align}
In $\hat{\rho}_{\rm out}$, the probability of a successful detection is given by the normalising factor,
\begin{align}
\mathcal{T} =  \frac{\eta^2}{8} + \frac{\eta(1-\eta)|\gamma|^2}{2} + \frac{(1-\eta)^2|\gamma|^4}{8} + \frac{2\mathcal{V}\eta^2 \mathcal{D}(1-\mathcal{D})}{(1+4\mathcal{D}(1-\mathcal{D}))^2}. 
\end{align}
Using the definition of operators as in Sec.~\ref{sec:StabExp},  the stabiliser expectation values on the input (qubits 1,2) and output states (qubits 1,4) are calculated to be:
\begin{align} \label{eq:stab_exps}
\rm{Input:}& \nonumber\\
\langle \hat{X}\hat{X} \rangle_{\rm in} &=\bra{\Phi^+} \hat{X}\hat{X} \ket{\Phi^+} &&= \frac{1}{(\eta+(1-\eta)|\gamma|^2)^2} ~ \frac{4k\eta(1-\eta)|\gamma|^2}{(1+k^2)} + \eta^2 + (1-\eta)^2|\gamma|^4,\\
\langle -\hat{Y}\hat{Y} \rangle_{\rm in} &= -\bra{\Phi^+} \hat{Y}\hat{Y} \ket{\Phi^+}  &&= \frac{\eta^2}{(\eta+(1-\eta)|\gamma|^2)^2} ~ \frac{(1-k^2 )}{(1+k^2)},\\
\langle \hat{Z}\hat{Z} \rangle_{\rm in} &= \bra{\Phi^+} \hat{Z}\hat{Z} \ket{\Phi^+}  &&= \frac{\eta^2}{(\eta+(1-\eta)|\gamma|^2)^2} ~ \frac{(1 - k^2)}{(1+k^2)}.\\
\rm{Output:}& \nonumber\\
\langle \hat{X}\hat{X} \rangle_{\rm out} &= \operatorname{Tr}{(\hat{\rho}_{\rm out} \hat{X}\hat{X} )} &&= \frac{1}{2\mathcal{T}} \left[ \mathcal{B}\left(\frac{2(1-\eta)|\gamma|^2k}{(1+k^2)} + \eta\right)^2 + (\mathcal{A}+\mathcal{C})\left(\frac{2\eta k}{(1+k^2)} + (1-\eta)|\gamma|^2\right)^2 \right],\\
\langle -\hat{Y}\hat{Y} \rangle_{\rm out} &= \operatorname{Tr}{(\hat{\rho}_{\rm out} -\hat{Y}\hat{Y} )} &&=  \frac{1}{8\mathcal{T}}\frac{(1-k^2)^2}{(1+k^2)^2} \mathcal{V}\eta^4,\\
\langle \hat{Z}\hat{Z} \rangle_{\rm out} &= \operatorname{Tr}{(\hat{\rho}_{\rm out} \hat{Z}\hat{Z} )} &&=  \frac{\eta^4}{8\mathcal{T}}  \frac{(1-k^2)^2}{(1+k^2)^2}.
\end{align}

Here we have normalised by the value of  $\bra{\Phi^+} \hat{I}\hat{I} \ket{\Phi^+} = \langle \hat{I}\hat{I}\rangle = {(\eta+(1-\eta)|\gamma|^2)^2}$.

The fusion map applied in terms of to the input and output stabilisers values is,
\begin{align} \label{eq:map_form}
P_{0f} &= \frac{1}{4}~ \left( \frac{\langle \hat{I}\hat{I}\rangle_{\rm out}}{\langle \hat{I}\hat{I}\rangle_{\rm in}^2} + \frac{\langle \hat{X}\hat{X}\rangle_{\rm out}}{\langle \hat{X}\hat{X}\rangle_{\rm in}^2} + \frac{\langle -\hat{Y}\hat{Y}\rangle_{\rm out}}{\langle -\hat{Y}\hat{Y}\rangle_{\rm in}^2} +\frac{\langle \hat{Z}\hat{Z}\rangle_{\rm out}}{\langle \hat{Z}\hat{Z}\rangle_{\rm in}^2}  \right),\\
P_{xf} &=  \frac{1}{4}~\left(\frac{\langle \hat{I}\hat{I}\rangle_{\rm out}}{\langle \hat{I}\hat{I}\rangle_{\rm in}^2} + \frac{\langle \hat{X}\hat{X}\rangle_{\rm out}}{\langle \hat{X}\hat{X}\rangle_{\rm in}^2} - \frac{\langle -\hat{Y}\hat{Y}\rangle_{\rm out}}{\langle -\hat{Y}\hat{Y}\rangle_{\rm in}^2} -\frac{\langle \hat{Z}\hat{Z}\rangle_{\rm out}}{\langle \hat{Z}\hat{Z}\rangle_{\rm in}^2}  \right),\\
P_{yf} &=  \frac{1}{4}~\left( \frac{\langle \hat{I}\hat{I}\rangle_{\rm out}}{\langle \hat{I}\hat{I}\rangle_{\rm in}^2} - \frac{\langle \hat{X}\hat{X}\rangle_{\rm out}}{\langle \hat{X}\hat{X}\rangle_{\rm in}^2} + \frac{\langle -\hat{Y}\hat{Y}\rangle_{\rm out}}{\langle -\hat{Y}\hat{Y}\rangle_{\rm in}^2} -\frac{\langle \hat{Z}\hat{Z}\rangle_{\rm out}}{\langle \hat{Z}\hat{Z}\rangle_{\rm in}^2}  \right),\\
P_{zf} &=  \frac{1}{4}~\left( \frac{\langle \hat{I}\hat{I}\rangle_{\rm out}}{\langle \hat{I}\hat{I}\rangle_{\rm in}^2} - \frac{\langle \hat{X}\hat{X}\rangle_{\rm out}}{\langle \hat{X}\hat{X}\rangle_{\rm in}^2} - \frac{\langle -\hat{Y}\hat{Y}\rangle_{\rm out}}{\langle -\hat{Y}\hat{Y}\rangle_{\rm in}^2} +\frac{\langle \hat{Z}\hat{Z}\rangle_{\rm out}}{\langle \hat{Z}\hat{Z}\rangle_{\rm in}^2}  \right).
\end{align}

We derive this by looking at at the input and output expressions for the ideal Bell state under a pauli channel. The input is
\begin{align}
    \hat{\rho}_{\rm in} =  p_o ~&\ket{\Phi^+}\bra{\Phi^+} 
    + p_x ~\hat{X}_1\ket{\Phi^+}\bra{\Phi^+}\hat{X}_1 \\
    + ~ p_y ~\hat{Y}_1&\ket{\Phi^+}\bra{\Phi^+}\hat{Y}_1 
    + p_z ~\hat{Z}_1\ket{\Phi^+}\bra{\Phi^+}\hat{Z}_1 .
\end{align}
We can consider the error is all on qubit 1, or equivalently under the action of stabiliser shift all error onto the first qubit. The input to the fusion measurement is then $\hat{\rho}_{\rm in ~1,2} \otimes \hat{\rho}_{\rm in ~3,4}$, where qubits 2,3 have no error, and are destructively measured out in the desired state with some probability. This gives us $\hat{\rho}_{\rm out'}$ of qubits 1,4 with errors on both qubits, where we can again shift all errors to qubit 1 :
\begin{align}
    \hat{\rho}_{\rm out'} =  ~~&(p_o^2 + p_x^2+p_y^2+p_z^2) ~\ket{\Phi^+}\bra{\Phi^+} \\
    + &(p_xp_o+p_op_x+p_yp_z +p_zp_y) ~\hat{X}_1\ket{\Phi^+}\bra{\Phi^+}\hat{X}_1 \\
    +  &(p_yp_o+p_op_y+p_xp_z +p_zp_x) ~\hat{Y}_1\ket{\Phi^+}\bra{\Phi^+}\hat{Y}_1 \\
    + &(p_zp_o+p_op_z+p_xp_y +p_yp_x) ~\hat{Z}_1\ket{\Phi^+}\bra{\Phi^+}\hat{Z}_1 .
\end{align}

The final output state is the density matrix through a Pauli error channel on one qubit :
\begin{align}
    \hat{\rho}_{\rm out } =  P_{of} ~&\hat{\rho}_{\rm out'} 
    + P_{xf} ~\hat{X}_1\hat{\rho}_{\rm out'}\hat{X}_1 \\
    + ~ P_{yf} ~\hat{Y}_1&\hat{\rho}_{\rm out'}\hat{Y}_1 
    + P_{zf} ~\hat{Z}_1\hat{\rho}_{\rm out'}\hat{Z}_1 .
\end{align}
Calculating the values of $\langle \hat{X}\hat{X} \rangle_{\rm out}, -\langle \hat{Y}\hat{Y} \rangle_{\rm out}, \langle \hat{Z}\hat{Z} \rangle_{\rm out}$, which are functions of $P_{of},P_{xf},P_{yf},P_{zf}$ and $p_o, p_x, p_y, p_z$ , and then expressing in terms of $\langle \hat{X}\hat{X} \rangle_{\rm in}, -\langle \hat{Y}\hat{Y} \rangle_{\rm in}, \langle \hat{Z}\hat{Z} \rangle_{\rm in}$, we get the expression in Eq. \ref{eq:map_form}.

Substituting the values \ref{eq:stab_exps} in \ref{eq:map_form},  we have the exact analytic probabilities, expanding to 1st order we get the desired resultant map listed in Sec.~\ref{tab:FusionMaps}. 

Repeating the same calculation for three-level emitters, the probability for a successful detection is,
\begin{equation}
    \mathcal{T} = \frac{3(1-\eta)^2|\gamma|^4}{2} + (1-\eta)|\gamma|^2\eta + \frac{\eta^2}{8}.
\end{equation}
And the input and output stabiliser values are:
\begin{align} \label{eq:stab_exps3}
\rm{Input:} &\nonumber\\
\langle \hat{X}\hat{X} \rangle_{\rm in} &=\bra{\Phi^+} \hat{X}\hat{X} \ket{\Phi^+} &&= \frac{\zeta^2\eta^2 + 4(1-\eta)^2|\gamma|^4}{(\eta + 2(1-\eta)|\gamma|^2)^2},\\
\langle -\hat{Y}\hat{Y}\rangle_{\rm in}  &= -\bra{\Phi^+} \hat{Y}\hat{Y} \ket{\Phi^+}  &&= \frac{\zeta^2\eta^2}{(\eta + 2(1-\eta)|\gamma|^2)^2},\\
\langle \hat{Z}\hat{Z}\rangle_{\rm in}  &= \bra{\Phi^+} \hat{Z}\hat{Z} \ket{\Phi^+}  &&= \frac{\eta^2}{(\eta + 2(1-\eta)|\gamma|^2)^2}.\\
\rm{Output:} &\nonumber\\
\langle \hat{X}\hat{X}\rangle_{\rm out}  &= \operatorname{Tr}{(\hat{\rho}_{\rm out} \hat{X}\hat{X} )} &&= \frac{4(1-\eta)^2|\gamma|^4}{(\eta + 2(1-\eta)|\gamma|^2)^2} + \frac{\zeta^4\eta^4}{8\mathcal{T}(\eta + 2(1-\eta)|\gamma|^2)^2}, \\
\langle -\hat{Y}\hat{Y}\rangle_{\rm out}  &= -\operatorname{Tr}{(\hat{\rho}_{\rm out} \hat{Y}\hat{Y} )} &&=  \frac{\zeta^4\eta^4}{8\mathcal{T}(\eta + 2(1-\eta)|\gamma|^2)^2},\\
\langle \hat{Z}\hat{Z}\rangle_{\rm out}  &= \operatorname{Tr}{(\hat{\rho}_{\rm out} \hat{Z}\hat{Z} )} &&=  \frac{\eta^4}{8\mathcal{T}(\eta + 2(1-\eta)|\gamma|^2)^2}.
\end{align}
Again, substituting the values \ref{eq:stab_exps3} in \ref{eq:map_form},  we have the exact analytic probabilities for the three-level emitter, and expanding to 1st order we get the resultant map listed in Sec.~\ref{tab:FusionMaps}. 

\section{Splitting Modes}\label{OpModeDeriv}

In the following section we construct Pauli operators for a dual rail configuration of single-photon qubits. The operators are expressed in terms of annihilation and creation operators of the photons of orthogonal temporal modes.

First we consider the single-photon identity operator $\hat{I}$ for a dual-rail qubit that defines the code space under number-resolving photon detection. The dual-rail code space is defined as having a single photon in rail 1 or 2 (denoted by subscripts 1 or 2) arriving within some time-cycle between $0$ and $\tau$. The measurement is insensitive to the arrival time of the photon and therefore we average over the time cycle. Taking this into account we write the identity operator as,
\begin{equation}\label{I_exp}
    \hat{I} = \int_0^\tau dt ~~~\hat{a}^{\dagger}_{1}(t)\ket{0}\bra{0}a_1(t) + \hat{a}^{\dagger}_{2}(t)\ket{0}\bra{0}\hat{a}_2(t).
\end{equation}
Here we denote the vacuum state as $\ket{0}$. If the photons are well localised in the time-cycle, we are justified in taking the limit $\tau \rightarrow \infty$. For all input states in the one-photon subspace, this operator leaves the states unaffected, and projects any multi-photon or zero photon states to 0.

For a complete set of orthonormal functions $f_n(t)$ ; $n \in \mathbb{N}$  we have   $\int_0^\infty dt f_n^*(t)f_m(t) = \delta_{n,m}$.
We can rewrite the photon ladder operators $\hat{a}^\dagger_{1/2}(t)/\hat{a}_{1/2}(t)$ in terms of a ``discretised'' set of operators, we refer to as `modes':
\begin{equation}\label{mode_eq}
\begin{split}
\hat{a}_{1/2}(t) &= \int_0^\infty {dt'}~{\delta(t-t')\hat{a}_{1/2}(t')} \\
&=\int_0^\infty  dt' ~~ \left[\sum_n^\infty f_n^*(t)f_n(t')\right]~~\hat{a}_{1/2}(t') \\
&=\sum_n^\infty f_n^*(t)\hat{a}_{1/2 , n},   ~~~~~~\text{  where } ~~~~  \hat{a}_{1/2 , n} = \int dt' f_n(t')\hat{a}_{1/2}(t'). \\
\end{split}
\end{equation}
The exact basis that is used here is irrelevant. The basis composed of modes of the largest photon occupation is obtained by diagonalising $G^{(1)}$ as described in Sec.~ \ref{sec:CohTomography}. Now we can express $\hat{I}$ in terms of these modes. Substituting \ref{mode_eq} in \ref{I_exp} :
\begin{equation}
\begin{split}
    \hat{I} &= \int_0^\tau dt ~~~\sum_{n,m}f_n(t)\hat{a}^{\dagger}_{1n}\ket{0}\bra{0}f_m^*(t)\hat{a}_{1m} + \sum_{n,m}f_n(t)\hat{a}^{\dagger}_{2n}\ket{0}\bra{0}f_m^*(t)\hat{a}_{2m} \\
    &= \sum_{n,m} \left(\hat{a}^{\dagger}_{1n}\ket{0}\bra{0}\hat{a}_{1m} ~ \int_0^\tau dt ~ f_n(t)f_m^*(t)\right) + \sum_{n,m} \left(\hat{a}^{\dagger}_{2n}\ket{0}\bra{0}\hat{a}_{2m} ~ \int_0^\tau dt ~ f_n(t)f_m^*(t)\right) \\
    &= \sum_{n,m} \left(\hat{a}^{\dagger}_{1n}\ket{0}\bra{0}\hat{a}_{1m} ~ \delta_{nm}\right) + \sum_{n,m} \left(\hat{a}^{\dagger}_{2n}\ket{0}\bra{0}\hat{a}_{2m} ~ \delta_{nm}\right) \\
    &= \sum_{n} \left(\hat{a}^{\dagger}_{1n}\ket{0}\bra{0}\hat{a}_{1n}  +  \hat{a}^{\dagger}_{2n}\ket{0}\bra{0}\hat{a}_{2n} ~ \right).
    \end{split}
\end{equation}
This can be intuitively understood as the identity operator acting on each single photon present in either of the rails, in any mode. 
For the constructed single photon $\hat{Z}$ operator, photons in rail 1 remain unaffected, and photons in rail 2 acquire a $\pi$ phase. Expressing this in terms of ladder operators we have:
\begin{align} \label{Z_exp}
    \hat{Z} = \int_0^\tau dt ~~~\hat{a}^{\dagger}_{1}(t)\ket{0}\bra{0}a_1(t) - \hat{a}^{\dagger}_{2}(t)\ket{0}\bra{0}a_2(t).
\end{align}
For the constructed $\hat{X}$ operator, the photons exchange rails, and for $\hat{Y}$, the photons exchange rails with rail 1 (2) acquiring a phase $(-) ~\pi/2$  :
\begin{align}
    \label{X_exp}
    \hat{X} = \int_0^\tau dt ~~~\hat{a}^{\dagger}_{1}(t)\ket{0}\bra{0}a_2(t) + \hat{a}^{\dagger}_{2}(t)\ket{0}\bra{0}a_1(t) ~,
\end{align}
\begin{align}
    \label{Y_exp}
    \hat{Y} = i\int_0^\tau dt ~~~\hat{a}^{\dagger}_{1}(t)\ket{0}\bra{0}a_2(t) - \hat{a}^{\dagger}_{2}(t)\ket{0}\bra{0}a_1(t) ~.
\end{align}
Similarly substituting \ref{mode_eq} in \ref{Z_exp}, \ref{X_exp} and \ref{Y_exp} we get the following expression for the operators:
\begin{align}
   \hat{Z} &= \sum_{n} \left(\hat{a}^{\dagger}_{1n}\ket{0}\bra{0}\hat{a}_{1n}  -  \hat{a}^{\dagger}_{2n}\ket{0}\bra{0}\hat{a}_{2n} ~ \right) ,\\
    \hat{X} &= \sum_{n} \left(\hat{a}^{\dagger}_{1n}\ket{0}\bra{0}\hat{a}_{2n}  +  \hat{a}^{\dagger}_{2n}\ket{0}\bra{0}\hat{a}_{1n} ~ \right) ,\\
     \hat{Y} &= i\sum_{n} \left(  \hat{a}^{\dagger}_{2n}\ket{0}\bra{0}\hat{a}_{1n} - \hat{a}^{\dagger}_{1n}\ket{0}\bra{0}\hat{a}_{2n}~ \right) .
\end{align}
We express $\hat{a}^{\dagger}_{1n}\ket{0}$ as the shorthand $\ket{1_n}$, i.e. one photon in mode $n$, which gives us Eqs.~\eqref{eq:SinglePhotonOps}-\eqref{eq:SinglePhotonOps4} in the main text.

\section{A Note on Different Measurement Outcomes} \label{ExtraFusion}

\renewcommand{\arraystretch}{2}
\setlength{\tabcolsep}{10pt}

In this Appendix we list the errors on remaining states after fusion for detection patterns other than $\ket{1100}/\ket{0011}$. 
These are calculated in the same way as specified in App.~\ref{FusionMapping}. One should note that the probabilities of each detection pattern being measured out are functions of the experimental parameters as well.

The other detection patters we have are $\ket{1010}/\ket{0101}$, which ideally leads to the resultant state $\ket{\Phi^-}$, or in the case of fusion failure, $\ket{2000}/\ket{0002}$ leads to the resultant state $\ket{01}$, and $\ket{0200}/\ket{0020}$  leads to the resultant state $\ket{10}$. Comparing errors on the input state, to errors on the output state we obtain the Pauli map in all these cases, and approximate them up to first order in terms of experimental parameters. 
Note that in the case of $\ket{\Phi^\pm}$, the Pauli map calculated is applied to one qubit, but in the case of fusion failure, we have calculated the bit-flip probability for each qubit.

For four-level emitters we have :

\begin{table}[H]
    \centering    \begin{tabular}{|c|c|c|c|}
\hline
         &  $\ket{\Phi^+}$&  $\ket{\Phi^-}$& Failure\\
         \hline
         Prob. of Detection&  $\frac{1}{4}(1-2\bar{\eta}+4\bar{\eta}|\gamma|^2+16\bar{\mathcal{D}}-144\bar{\mathcal{D}}^2)$&  $\frac{1}{4}(1-2\bar{\eta}+4\bar{\eta}|\gamma|^2-16\bar{\mathcal{D}}+ 144\bar{\mathcal{D}}^2)$& $\frac{1}{2}(1-2\bar{\eta}+4\bar{\eta}|\gamma|^2)$\\
         $p_o$&  $1-\frac{3}{2}\bar{\eta}|\gamma|^2-\frac{\bar{\mathcal{V}}}{2} -8\bar{\mathcal{D}}+200\bar{\mathcal{D}}^2$&  $1-\frac{3}{2}\bar{\eta}|\gamma|^2-\frac{\bar{\mathcal{V}}}{2}+8\bar{\mathcal{D}}+56\bar{\mathcal{D}}^2$& $1-\frac{3}{2}\bar{\eta}|\gamma|^2-4\bar{\mathcal{D}}-12\bar{\mathcal{D}}^2$\\
         $p_x$&  $\frac{1}{2}\bar{\eta}|\gamma|^2+8\bar{\mathcal{D}}-200\bar{\mathcal{D}}^2$&  $\frac{1}{2}\bar{\eta}|\gamma|^2-8\bar{\mathcal{D}}-56\bar{\mathcal{D}}^2$& $\frac{3}{2}\bar{\eta}|\gamma|^2+4\bar{\mathcal{D}}+12\bar{\mathcal{D}}^2$\\
         $p_y$&  $\frac{1}{2}\bar{\eta}|\gamma|^2$&  $\frac{1}{2}\bar{\eta}|\gamma|^2$& -\\
 $p_z$& $\frac{1}{2}\bar{\eta}|\gamma|^2+\frac{\bar{\mathcal{V}}}{2}$& $\frac{1}{2}\bar{\eta}|\gamma|^2+\frac{\bar{\mathcal{V}}}{2}$&-\\
   \hline
    \end{tabular}
    \caption{Probability of application of $\hat{I}, \hat{X}, \hat{Y} ~\text{and} ~\hat{Z}$ respectively for the Pauli map applied due to fusion measurements, for different resultant states for four-level emitters.}
    \label{tab:allfusion4lev}
\end{table}

And for three-level emitters:

\begin{table}[H]
    \centering
    \begin{tabular}{|c|c|c|c|}
    \hline
         &  $\ket{\Phi^+}$&  $\ket{\Phi^-}$& Failure\\
         \hline
         Prob. of Detection&  $\frac{1}{4}(1-2\bar{\eta}+8\bar{\eta}|\gamma|^2)$&  $\frac{1}{4}(1-2\bar{\eta}+8\bar{\eta}|\gamma|^2)$& $\frac{1}{4}(1-2\bar{\eta}+8\bar{\eta}|\gamma|^2)$\\
         $p_o$&  $1-3\bar{\eta}|\gamma|^2$&  $1-3\bar{\eta}|\gamma|^2$& $1-3\bar{\eta}|\gamma|^2$\\
         $p_x$&  $\bar{\eta}|\gamma|^2$&  $\bar{\eta}|\gamma|^2$& $3\bar{\eta}|\gamma|^2$\\
         $p_y$&  $\bar{\eta}|\gamma|^2$&  $\bar{\eta}|\gamma|^2$& -\\
 $p_z$& $\bar{\eta}|\gamma|^2$& $\bar{\eta}|\gamma|^2$&-\\
 \hline
    \end{tabular}
    \caption{Probability of application of $\hat{I}, \hat{X}, \hat{Y} ~\text{and} ~\hat{Z}$ respectively for the Pauli map applied due to fusion measurements, for different resultant states for three-level emitters.}
    \label{tab:allfusion3lev}
\end{table}

For state emission, i.e. the chain, GHZ and branched states produced in Sec. \ref{sec:Wavefuncions}, we would like to note measuring out the emitter in the positive or negative eigenstate results in different mappings or errors on the qubits. In the main text, we've considered positive measurement outcomes. For completeness the errors on measuring the negative eigenvalue remain the same up to first order.  
\section{Complete Stabiliser and Error expressions} \label{StabVals}

\renewcommand{\arraystretch}{2}
\setlength{\tabcolsep}{12pt}

In the following Appendix we list calculate stabiliser expectation values for chain states and GHZ states, as well as the dependence of Pauli errors on these stabiliser values. As described in Sec.~\ref{sec:StabExp}, any stabiliser value can be calculated, and then used to infer Pauli errors. Here however we have used low weight, and localised stabilisers: stabiliser generators, nearest neighbour, and next nearest products of the generators.

For chain states, we have the following expressions for stabiliser expectation values in the bulk:
\begin{table}[H]
    \centering
    \begin{tabular}{|c|c|c|}
    \hline
         &  Three-Level &  Four-Level (Even $N$)\\
         \hline
         $\langle \hat{Z}_{i-1}\hat{X}_i\hat{Z}_{i+1}\rangle$&  $\zeta\eta^3 /(\eta+2(1-\eta)|\gamma|^2)^3$&  $(1-k^2)\eta^2(\eta + k(1-\eta)|\gamma|^2) /(\eta+(1-\eta)|\gamma|^2)^3$\\
         $\langle \hat{Z}_{i-2}\hat{X}_{i-1}\hat{I}_i\hat{X}_{i+1}\hat{Z}_{i+1}\rangle$&  $\zeta^2\eta^4 /(\eta+2(1-\eta)|\gamma|^2)^4$&  $(1-k^2)\eta^2(\eta + k(1-\eta)|\gamma|^2)^2 /(\eta+(1-\eta)|\gamma|^2)^4$\\
         $\langle \hat{Z}_{i-1}\hat{Y}_{i}\hat{Y}_{i+1}\hat{Z}_{i+2}\rangle$&  $\zeta^2\eta^4 /(\eta+2(1-\eta)|\gamma|^2)^4$&  $(1-k^2)^2\eta^4 /(\eta+(1-\eta)|\gamma|^2)^4$\\
         \hline
    \end{tabular}
    \caption{Stabiliser expectation values in the bulk of an $N$-long chain emitted from three or four-level emitters, in terms of experimental parameters}
    \label{tab:apStabValTab1}
\end{table}
where $k$ is a function of the birefringence, namely $k=2\sqrt{\mathcal{D}(1-\mathcal{D})}$.
Relating Pauli errors to stabiliser expectation values using \ref{eq:stabexp_to_pauli} we have:

\begin{align} \label{eq:chainPaulifromStab}
    p_x &= \frac{1}{4}\left( 1 + \frac{\langle \hat{Z}\hat{X}\hat{I}\hat{X}\hat{Z}\rangle}{\langle \hat{Z}\hat{X}\hat{Z}\rangle} - \frac{\sqrt{\langle \hat{Z}\hat{X}\hat{I}\hat{X}\hat{Z}\rangle \langle \hat{Z}\hat{Y}\hat{Y}\hat{Z}\rangle}}{\langle \hat{Z}\hat{X}\hat{Z}\rangle}- \frac{\langle \hat{Z}\hat{X}\hat{Z}\rangle}{\sqrt{\langle \hat{Z}\hat{X}\hat{I}\hat{X}\hat{Z}\rangle}}\right), \\
    p_y &= \frac{1}{4}\left( 1 - \frac{\langle \hat{Z}\hat{X}\hat{I}\hat{X}\hat{Z}\rangle}{\langle \hat{Z}\hat{X}\hat{Z}\rangle} + \frac{\sqrt{\langle \hat{Z}\hat{X}\hat{I}\hat{X}\hat{Z}\rangle \langle \hat{Z}\hat{Y}\hat{Y}\hat{Z}\rangle}}{\langle \hat{Z}\hat{X}\hat{Z}\rangle}- \frac{\langle \hat{Z}\hat{X}\hat{Z}\rangle}{\sqrt{\langle \hat{Z}\hat{X}\hat{I}\hat{X}\hat{Z}\rangle}}\right), \\
    p_z &= \frac{1}{4}\left( 1 - \frac{\langle \hat{Z}\hat{X}\hat{I}\hat{X}\hat{Z}\rangle}{\langle \hat{Z}\hat{X}\hat{Z}\rangle} - \frac{\sqrt{\langle \hat{Z}\hat{X}\hat{I}\hat{X}\hat{Z}\rangle \langle \hat{Z}\hat{Y}\hat{Y}\hat{Z}\rangle}}{\langle \hat{Z}\hat{X}\hat{Z}\rangle}+ \frac{\langle \hat{Z}\hat{X}\hat{Z}\rangle}{\sqrt{\langle \hat{Z}\hat{X}\hat{I}\hat{X}\hat{Z}\rangle}}\right). \\
\end{align}
Where we have neglected the subscripts of the operators, but are the same as in Table. \ref{tab:apStabValTab1}.

Substituting values from in \ref{eq:chainPaulifromStab} gives us:

\begin{table}[H]
    \centering
    \begin{tabular}{|c|c|c|}
    \hline
         &  Three-Level&  Four-Level (Even $N$)\\
         \hline
         $p_x$&  $\frac{(1-\eta)|\gamma|^2}{2(2(1-\eta)|\gamma|^2 +\eta)}$&  $\frac{4\eta(1-\mathcal{D}) + (1-\eta)|\gamma|^2(1+2\sqrt{\mathcal{D}(1-\mathcal{D})})}{4((1-\eta)|\gamma|^2 +\eta)}$\\
         $p_y$&  $\frac{(1-\eta)|\gamma|^2}{2(2(1-\eta)|\gamma|^2 +\eta)}$&  $\frac{(1-\eta)|\gamma|^2(1-2\sqrt{\mathcal{D}(1-\mathcal{D})})}{4((1-\eta)|\gamma|^2 +\eta)}$\\
         $p_z$&  $\frac{\eta(1-\zeta) + (1-\eta)|\gamma|^2}{2(2(1-\eta)|\gamma|^2 +\eta)}$&  $\frac{(1-\eta)|\gamma|^2(1-2\sqrt{\mathcal{D}(1-\mathcal{D})})}{4((1-\eta)|\gamma|^2 +\eta)}$\\
         \hline
    \end{tabular}
    \caption{Values of Pauli error rates on each qubit in the bulk of an $N$-long chain emitted from three or four-level emitters, in terms of experimental parameters}
    \label{tab:apStabValTab2}
\end{table}

For GHZ states we consider the following stabiliser expectation values:

\begin{table}[H]
    \centering
    \begin{tabular}{|c|c|c|}
    \hline
         &  Three-Level &  Four-Level\\
         \hline
         $\langle Z_iZ_j\rangle ~~\forall ~ i,j \in [1,N], i\neq j$&  $\frac{\eta^2}{(2(1-\eta)|\gamma|^2  +\eta)^2}$&  $\frac{\eta^2(2\mathcal{D}-1)^2}{((1-\eta)|\gamma|^2  +\eta)^2 (1+k^N)}$\\
         $\langle \hat{X}^{\otimes N}\rangle$&  $\frac{(\eta{\zeta})^N + (2(1-\eta)|\gamma|^2 )^N}{(2(1-\eta)|\gamma|^2  +\eta)^N}$&  $\frac{(\eta + k(1-\eta)|\gamma|^2)^N + ((1-\eta)|\gamma|^2+k\eta)^N}{((1-\eta)|\gamma|^2 +\eta)^N(1+k^N)}$\\
         $\langle \hat{X}^{\otimes N-2}\hat{Y}_i\hat{Y}_j\rangle ~~\forall ~ i,j \in [1,N], i\neq j$&  $\frac{(\eta{\zeta})^N}{(2(1-\eta)|\gamma|^2  +\eta)^N}$&  $\frac{\eta^2(\eta + k(1-\eta)|\gamma|^2)^{N-2}(2\mathcal{D}-1)^2}{((1-\eta)|\gamma|^2 +\eta)^N(1+k^N)}$\\
         \hline
    \end{tabular}
    \caption{Stabiliser expectation values for an $N$-qubit GHZ state emitted from three or four-level emitters, in terms of experimental parameters}
    \label{tab:apStabValTab3}
\end{table}

where the Pauli errors for the central qubit are given by:

\begin{align}
    p_x &= \frac{1}{4}\left( 1 + 
\langle \hat{X}^{\otimes N}\rangle
\left(\frac{1+\sqrt{\langle\hat{Z_i}\hat{Z_j}\rangle}}{1+\sqrt{\frac{\langle\hat{X}^{\otimes N}\rangle}{\langle\hat{X}^{\otimes N-2}\hat{Y_i}\hat{Y_j}\rangle}}}\right)^{1-N}
-\sqrt{\langle \hat{X}^{\otimes N}\rangle  \langle \hat{X}^{\otimes N-2}\hat{Y}_i\hat{Y}_j\rangle}\left(\frac{1+\sqrt{\langle\hat{Z_i}\hat{Z_j}\rangle}}{1+\sqrt{\frac{\langle\hat{X}^{\otimes N}\rangle}{\langle\hat{X}^{\otimes N-2}\hat{Y_i}\hat{Y_j}\rangle}}}\right)^{1-N}
- \sqrt{\langle \hat{Z}_i\hat{Z}_j\rangle}\right), \\
    p_y &= \frac{1}{4}\left( 1 - 
\langle \hat{X}^{\otimes N}\rangle
\left(\frac{1+\sqrt{\langle\hat{Z_i}\hat{Z_j}\rangle}}{1+\sqrt{\frac{\langle\hat{X}^{\otimes N}\rangle}{\langle\hat{X}^{\otimes N-2}\hat{Y_i}\hat{Y_j}\rangle}}}\right)^{1-N}
+\sqrt{\langle \hat{X}^{\otimes N}\rangle  \langle \hat{X}^{\otimes N-2}\hat{Y}_i\hat{Y}_j\rangle}\left(\frac{1+\sqrt{\langle\hat{Z_i}\hat{Z_j}\rangle}}{1+\sqrt{\frac{\langle\hat{X}^{\otimes N}\rangle}{\langle\hat{X}^{\otimes N-2}\hat{Y_i}\hat{Y_j}\rangle}}}\right)^{1-N}
- \sqrt{\langle \hat{Z}_i\hat{Z}_j\rangle}\right), \\
    p_z &= \frac{1}{4}\left( 1 -
\langle \hat{X}^{\otimes N}\rangle
\left(\frac{1+\sqrt{\langle\hat{Z_i}\hat{Z_j}\rangle}}{1+\sqrt{\frac{\langle\hat{X}^{\otimes N}\rangle}{\langle\hat{X}^{\otimes N-2}\hat{Y_i}\hat{Y_j}\rangle}}}\right)^{1-N}
-\sqrt{\langle \hat{X}^{\otimes N}\rangle  \langle \hat{X}^{\otimes N-2}\hat{Y}_i\hat{Y}_j\rangle}\left(\frac{1+\sqrt{\langle\hat{Z_i}\hat{Z_j}\rangle}}{1+\sqrt{\frac{\langle\hat{X}^{\otimes N}\rangle}{\langle\hat{X}^{\otimes N-2}\hat{Y_i}\hat{Y_j}\rangle}}}\right)^{1-N}
+ \sqrt{\langle \hat{Z}_i\hat{Z}_j\rangle}\right).
\end{align}

And for the boundary qubits:

\begin{align}
    p_x &= \frac{1}{2}\left( 1 - \frac{1+\sqrt{\langle\hat{Z_i}\hat{Z_j}\rangle}}{1+\sqrt{\frac{\langle\hat{X}^{\otimes N}\rangle}{\langle\hat{X}^{\otimes N-2}\hat{Y_i}\hat{Y_j}\rangle}}} \right), \\
    p_y &= \frac{1}{2}\left( 1 - \frac{1+\sqrt{\langle\hat{Z_i}\hat{Z_j}\rangle}}{1+\sqrt{\frac{\langle\hat{X}^{\otimes N-2}\hat{Y_i}\hat{Y_j}\rangle}{\langle\hat{X}^{\otimes N}\rangle}}} \right),\\
    p_z &= 0 .
\end{align}
The analytical expression of these Pauli errors can be calculated by substituting values from Table. \ref{tab:apStabValTab3}, and is shown up to first order approximation in Table \ref{tab:3_4_Level}.
\section{Calculating $\beta_n$}\label{BetaDeriv}

In the following Appendix we describe how  $|\beta_n| = |\langle \phi_n| \phi_s \rangle|$ can be obtained experimentally. Consider the emitter starting out in an equal superposition of the ground states, and performing a cycle of excitation and emission,
\begin{align}
&\ket{\psi} =&& \frac{\ket{g}+\ket{s}}{\sqrt{2}}, \\
&\xrightarrow[\rm{\& ~emmiting}]{\rm{exciting}} &&\frac{1}{\sqrt{2}} \left( \ket{g}\ket{\phi_n}\ket{1_n^{\rm ph}} + \ket{s}\ket{\phi_s}\ket{0} \right) ~~ \ket{\gamma}, \\
&\xrightarrow[\rm{pulse}]{\Pi} &&\frac{1}{\sqrt{2}} \left( \ket{s}\ket{\phi_n}\ket{1_n^{\rm ph}} + \ket{g}\ket{\phi_s}\ket{0} \right) ~~ \ket{\gamma}, \\
&\xrightarrow[\rm{\& ~emmiting}]{\rm{exciting}}& &\frac{1}{\sqrt{2}} \left( \ket{s}\ket{\phi_s\phi_n}\ket{01_n^{\rm ph}} + \ket{g}\ket{\phi_n\phi_s}\ket{1^{ph}_n0} \right) ~~ \ket{\gamma\gamma}, \\
&\xrightarrow[\rm{pulse}]{\Pi/2} &&\frac{1}{\sqrt{2}} \left( 
\frac{\ket{g}+\ket{s}}{\sqrt{2}}\ket{\phi_s\phi_n}\ket{01_n^{\rm ph}} + \frac{\ket{g}-\ket{s}}{\sqrt{2}}\ket{\phi_n\phi_s}\ket{1^{ph}_n0} \right) ~~ \ket{\gamma\gamma}.
\end{align}
Here, we have used the shorthand $\ket{1^{\rm ph}_n} = \sqrt{\eta}\ket{1_n}_{\rm w}\ket{0}_{\rm l} + \sqrt{1-\eta}\ket{0}_{\rm w}\ket{1_n}_{\rm l}$, and $\ket{1_n}$ is a photon in the $n^{th}$ mode as defined in Appendix~\ref{OpModeDeriv}, with probability amplitude $\alpha_n$.
Measuring out the emitter in the $\ket{g}$ state, we have an equal superposition of the early and late photons. Note also, that we assume there is a time difference of $\tau$  between when early and late photons are emitted,
\begin{align}
    \ket{\psi} = \frac{1}{\sqrt{2}} \sum_n \alpha_n  \left[\ket{\phi_s\phi_n}\ket{0 1^{\rm ph}_n} + \ket{\phi_n\phi_s}\ket{ 1^{\rm ph}_n 0} \right] \ket{\gamma\gamma}_{\rm wc}.
\end{align}
To calculate the first-order correlator for this state, $G^{(1)}(t_1,t_2) = \langle \hat{a}^\dagger(t_1)\hat{a}(t_2)\rangle$ , we calculate the action of the annihilation operators on the state $\ket{\psi}$. Here we consider the specific case where  $t_2 > \tau$ and $t_1 < \tau$  so as to act separately on the early and late photons. The state is rearranged to consider all the states the annihilation operator can act on together, which includes the coherent states in the waveguide.

\begin{align}\label{eq:a1}
    \hat{a}(t_1) \ket{\psi} =&\sum_n \alpha_n ~ ~\ket{\phi_s\phi_n}\ket{0} \ket{\gamma}_{\rm wc}(\hat{a}(t_1)\ket{ 1^{\rm ph}_n}\ket{\gamma}_{\rm wc}) + \ket{\phi_n\phi_s}\ket{ 1^{\rm ph}_n}\ket{\gamma}_{\rm wc} (\hat{a}(t_1)\ket{0}\ket{\gamma}_{\rm wc})\\
\Rightarrow  &\sum_n \alpha_n ~~ \ket{\phi_s\phi_n}\ket{0} \ket{\gamma}_{\rm wc}(\sqrt{\eta}\ket{0}_{\rm l}\hat{a}(t_1)\ket{1_n}_{\rm w}\ket{\gamma}_{\rm wc} + \sqrt{1-\eta}\ket{1_n}_{\rm l}\hat{a}(t_1)\ket{0}_{\rm w}\ket{\gamma}_{\rm wc}) \nonumber\\
& ~~~~~~~~~~~~~~~+ 
    \gamma c(t_1)\ket{\phi_n\phi_s}\ket{ 1^{\rm ph}_n 0}\ket{\gamma\gamma}_{\rm wc}  \nonumber\\
        \Rightarrow  &\sum_n \alpha_n ~~ \ket{\phi_s\phi_n}\ket{0} \ket{\gamma}_{\rm wc}\Huge{(}\sqrt{\eta}\ket{0}_{\rm l}\left(\int dt f_n^*(t) \left([\hat{a}(t_1) \hat{a}^\dagger(t)] + \hat{a}^\dagger(t)\hat{a}(t_1)\right) \right)\ket{0}_{\rm w}\ket{\gamma}_{\rm wc} 
        \nonumber \\
               &~~~~~~~~~~~~~~~+ \gamma c(t_1)\sqrt{1-\eta}\ket{1_n}_{\rm l}\ket{0}_{\rm w}\ket{\gamma}_{\rm wc} \Huge{)}
 +\gamma c(t_1)\ket{\phi_n\phi_s}\ket{ 1^{\rm ph}_n 0}\ket{\gamma\gamma}_{\rm wc}  \nonumber\\
                \Rightarrow  &\sum_n \alpha_n ~~ \sqrt{\eta}f_n^*(t_1)\ket{\phi_s\phi_n}\ket{\gamma\gamma}_{\rm wc}
            + \gamma c(t_1)\ket{\phi_s\phi_n}\ket{01_n^{\rm ph}}\ket{\gamma\gamma}_{\rm wc} + 
    \gamma c(t_1)\ket{\phi_n\phi_s}\ket{ 1^{\rm ph}_n 0}\ket{\gamma\gamma}_{\rm wc}  \nonumber\\
    \Rightarrow  &~~~~\gamma c(t_1)\ket{\psi} + \sqrt{\eta} \left(\sum_n \alpha_n f_n^*(t_1)\ket{\phi_s\phi_n}\right) \ket{\gamma\gamma}_{\rm wc} .
\end{align}
Here $c(t)$ is the normalised temporal mode of the laser in the waveguide, which can be obtained by measuring the laser light leaked into the waveguide in the absence of any emitted photon. $f_n^*(t_1)$ is the temporal mode of the $n^{th}$ temporal mode of the emitted photon we calculate in \ref{sec:CohTomography}  along with the $\alpha_n$'s .
For time $t_2 >\tau$  the annihilation operator acts in a similar way, but captures a different state of the emitters phonon environment :

\begin{align}\label{eq:a2}
    \hat{a}(t_2) \ket{\psi} =& \gamma c(t_2)\ket{\psi} + \sqrt{\eta} \left(\sum_n \alpha_n f_n^*(t_2)\ket{\phi_n\phi_s}\right) \ket{\gamma\gamma}_{\rm wc}. 
\end{align}

With the expressions \ref{eq:a1} and \ref{eq:a2} we have an expression of the first order coherence function :
\begin{align}
G^{(1)}(t_1,t_2) = |\gamma|^2c(t_2)c^*(t_1) + \eta\sum_{n,m}\alpha_n\alpha_m^*\beta_n^*\beta_m f_n^*(t_2) f_m(t_1).
\nonumber
\end{align}

Where $\beta_n = \braket{\phi_n}{\phi_s}$ . Using the orthonormality of the functions $\int dt f_n^*(t) f_m(t)=\delta_{n,m}$ we can simplify the right side of the equation to eliminate the summation.

\begin{align}
\iint dt_1 dt_2  ~~ f_i(t_2)\left( {G^{(1)}(t_1,t_2) - |\gamma|^2 c^*(t_1)c(t_2)} \right) f^*_{i}(t_1)&= \eta \sum_{n,m}\iint dt_1 dt_2 ~~\alpha_n\alpha_m^*\beta_n^*\beta_m f_i(t_2)f_n^*(t_2) f_m(t_1)f^*_i(t_1) \\
&= \eta \sum_{n,m}\alpha_n\alpha_m^*\beta_n^*\beta_m \delta_{n,i} \delta_{m,i}\nonumber \\
&= \eta |\alpha_i|^2|\beta_i|^2 . \nonumber\\
\end{align}
This rearranged gives us the value of any $|\beta_i|^2$ in terms of the measured coherence function $G^{(1)}$, temporal mode functions of the emitted photon and coherent laser, as well as average photon number of the desired mode and coherent laser and efficiency $\eta$ which were calculated in \ref{sec:CohTomography} :

\begin{align}
    |\beta_i|^2 = \frac{1}{\eta ~ |\alpha_i|^2} 
    \iint dt_1 dt_2  ~~ f_i(t_2) G^{(1)}(t_1,t_2)f_i^*(t_1)
    - |\gamma|^2 ~ \left| {\int dt ~c(t)f_i(t)}\right|^2 .
\end{align}

\end{document}